\documentclass[11pt]{article}
\usepackage{amsbsy}

\begin{document}

\rightline{ CU-TP-989, ITP-99/32}
\bigskip
\bigskip
\centerline{\LARGE \sc
Higgs Bundles and Four Manifolds}
\bigskip
\centerline{\sc Jae-Suk Park}   
\medskip
\centerline{\small Department of Physics, Columbia University}
\centerline{\small 538 West 120th Street, New York, 
N.Y.~10027, U.S.A\footnote{The present address}}
\centerline{(jspark@phys.columbia.edu)}
\centerline{\small and}
\centerline{\small Institute for Theoretical Physics, University of Amsterdam}
\centerline{\small Valckenierstraat 65, 1018 XE Amsterdam,
The Netherlands}

\bigskip

\begin{abstract}
It is known that the Seiberg-Witten invariants, derived
from supersymmetric Yang-Mill theories in four-dimensions,
do not distinguish smooth structure of certain non-simply-connected
four manifolds.
We propose generalizations of Donaldson-Witten and Vafa-Witten
theories on a K\"{a}hler manifold based on Higgs Bundles.
We showed, in particular, that the partition function
of our generalized Vafa-Witten theory can be written
as the sum of contributions our generalized Donaldson-Witten
invariants and generalized Seiberg-Witten invariants.
The resulting generalized Seiberg-Witten invariants might have,
conjecturally, information on smooth structure
beyond the original Seiberg-Witten invariants 
for non-simply-connected case.

\end{abstract}

%
%
\catcode`\@=11 

%
\global\newcount\secno \global\secno=0
\global\newcount\meqno \global\meqno=1

\def\newsec#1{\global\advance\secno by1
\global\subsecno=0\eqnres@t
\section{ #1}
}
\def\eqnres@t{\xdef\secsym{\the\secno.}\global\meqno=1}
\def\sequentialequations{\def\eqnres@t{\bigbreak}}\xdef\secsym{}
\global\newcount\subsecno \global\subsecno=0
\def\subsec#1{\global\advance\subsecno by1
\subsection{#1}}

\def\draftmode{\message{ DRAFTMODE }
\writelabels
 {\count255=\time\divide\count255 by 60 \xdef\hourmin{\number\count255}
  \multiply\count255 by-60\advance\count255 by\time
  \xdef\hourmin{\hourmin:\ifnum\count255<10 0\fi\the\count255}}}
\def\nolabels{\def\wrlabeL##1{}\def\eqlabeL##1{}\def\reflabeL##1{}}
\def\writelabels{\def\wrlabeL##1{\leavevmode\vadjust{\rlap{\smash%
{\line{{\escapechar=` \hfill\rlap{\tt\hskip.03in\string##1}}}}}}}%
\def\eqlabeL##1{{\escapechar-1\rlap{\tt\hskip.05in\string##1}}}%
\def\reflabeL##1{\noexpand\llap{\noexpand\sevenrm\string\string\string##1}}}
\nolabels

\def\eqn#1#2{
\xdef #1{(\secsym\the\meqno)}
\global\advance\meqno by1
$$#2\eqno#1\eqlabeL#1
$$}

\def\eqalign#1{\null\,\vcenter{\openup\jot\m@th
  \ialign{\strut\hfil$\displaystyle{##}$&$\displaystyle{{}##}$\hfil
      \crcr#1\crcr}}\,}

\def\foot#1{\footnote{#1}} 

%
\global\newcount\refno \global\refno=1
\newwrite\rfile
\def\ref{[\the\refno]\nref}
\def\nref#1{\xdef#1{[\the\refno]}
\ifnum\refno=1\immediate\openout\rfile=refs.tmp\fi
\global\advance\refno by1\chardef\wfile=\rfile\immediate
\write\rfile{\noexpand\bibitem{#1}}\findarg}
\def\findarg#1#{\begingroup\obeylines\newlinechar=`\^^M\pass@rg}
{\obeylines\gdef\pass@rg#1{\writ@line\relax #1^^M\hbox{}^^M}%
\gdef\writ@line#1^^M{\expandafter\toks0\expandafter{\striprel@x #1}%
\edef\next{\the\toks0}\ifx\next\em@rk\let\next=\endgroup\else\ifx\next\empty%
\else\immediate\write\wfile{\the\toks0}\fi\let\next=\writ@line\fi\next\relax}}
\def\striprel@x#1{} \def\em@rk{\hbox{}} 
\def\lref{\begingroup\obeylines\lr@f}
\def\lr@f#1#2{\gdef#1{\ref#1{#2}}\endgroup
}
\def\semi{;\hfil\break}
\def\addref#1{\immediate\write\rfile{\noexpand\item{}#1}} 
\def\listrefs{
}
\def\startrefs#1{\immediate\openout\rfile=refs.tmp\refno=#1}
\def\xref{\expandafter\xr@f}\def\xr@f[#1]{#1}
\def\refs#1{\count255=1[\r@fs #1{\hbox{}}]}
\def\r@fs#1{\ifx\und@fined#1\message{reflabel \string#1 is undefined.}%
\nref#1{need to supply reference \string#1.}\fi%
\vphantom{\hphantom{#1}}\edef\next{#1}\ifx\next\em@rk\def\next{}%
\else\ifx\next#1\ifodd\count255\relax\xref#1\count255=0\fi%
\else#1\count255=1\fi\let\next=\r@fs\fi\next}
\newwrite\lfile
{\escapechar-1\xdef\pctsign{\string\%}\xdef\leftbracket{\string\{}
\xdef\rightbracket{\string\}}\xdef\numbersign{\string\#}}
\def\writedefs{\immediate\openout\lfile=labeldefs.tmp \def\writedef##1{%
\immediate\write\lfile{\string\def\string##1\rightbracket}}}
\def\writestop{\def\writestoppt{\immediate\write\lfile{\string\pageno%
\the\pageno\string\startrefs\leftbracket\the\refno\rightbracket%
\string\def\string\secsym\leftbracket\secsym\rightbracket%
\string\secno\the\secno\string\meqno\the\meqno}\immediate\closeout\lfile}}

\catcode`\@=12 
%

%
\def\noblackbox{\overfullrule=0pt}
\hyphenation{anom-aly anom-alies coun-ter-term coun-ter-terms
}
\def\inv{^{\raise.15ex\hbox{${\scriptscriptstyle -}$}\kern-.05em 1}}
\def\dup{^{\vphantom{1}}}
\def\Dsl{\,\raise.15ex\hbox{/}\mkern-13.5mu D} 
\def\dsl{\raise.15ex\hbox{/}\kern-.57em\partial}
\def\del{\partial}
\def\Psl{\dsl}
\def\tr{{\rm tr}} \def\Tr{{\rm Tr}}
\font\bigit=cmti10 scaled \magstep1
\def\biglie{\hbox{\bigit\$}} 
\def\lspace{\ifx\answ\bigans{}\else\qquad\fi}
\def\lbspace{\ifx\answ\bigans{}\else\hskip-.2in\fi} 
\def\boxeqn#1{\vcenter{\vbox{\hrule\hbox{\vrule\kern3pt\vbox{\kern3pt
	\hbox{${\displaystyle #1}$}\kern3pt}\kern3pt\vrule}\hrule}}}
\def\tilde{\widetilde} \def\bar{\overline} \def\hat{\widehat}
%
\def\CAG{{\cal A/\cal G}} \def\CO{{\cal O}} 
\def\CA{{\cal A}} \def\CC{{\cal C}} \def\CF{{\cal F}} \def\CG{{\cal G}} 
\def\CL{{\cal L}} \def\CH{{\cal H}} \def\CI{{\cal I}} \def\CU{{\cal U}}
\def\CB{{\cal B}} \def\CR{{\cal R}} \def\CD{{\cal D}} \def\CT{{\cal T}}
\def\e#1{{\rm e}^{^{\textstyle#1}}}
\def\grad#1{\,\nabla\!_{{#1}}\,}
\def\gradgrad#1#2{\,\nabla\!_{{#1}}\nabla\!_{{#2}}\,}
\def\ph{\varphi}
\def\psibar{\overline\psi}
\def\om#1#2{\omega^{#1}{}_{#2}}
\def\vev#1{\langle #1 \rangle}
\def\lform{\hbox{$\sqcup$}\llap{\hbox{$\sqcap$}}}
\def\darr#1{\raise1.5ex\hbox{$\leftrightarrow$}\mkern-16.5mu #1}
\def\lie{\hbox{\it\$}} 
\def\ha{{1\over2}}
\def\half{{\textstyle{1\over2}}} 
\def\roughly#1{\raise.3ex\hbox{$#1$\kern-.75em\lower1ex\hbox{$\sim$}}}

%

\font\teneufm=eufm10
\font\seveneufm=eufm7
\font\fiveeufm=eufm5
\newfam\eufmfam
\textfont\eufmfam=\teneufm
\scriptfont\eufmfam=\seveneufm
\scriptscriptfont\eufmfam=\fiveeufm
\def\eufm#1{{\fam\eufmfam\relax#1}}

\font\teneurm=eurm10
\font\seveneurm=eurm7
\font\fiveeurm=eurm5
\newfam\eurmfam
\textfont\eurmfam=\teneurm
\scriptfont\eurmfam=\seveneurm
\scriptscriptfont\eurmfam=\fiveeurm
\def\eurm#1{{\fam\eurmfam\relax#1}}

\font\tenmsx=msam10
\font\sevenmsx=msam7
\font\fivemsx=msam5
\font\tenmsy=msbm10
\font\sevenmsy=msbm7
\font\fivemsy=msbm5
\newfam\msafam
\newfam\msbfam
\textfont\msafam=\tenmsx  \scriptfont\msafam=\sevenmsx
  \scriptscriptfont\msafam=\fivemsx
\textfont\msbfam=\tenmsy  \scriptfont\msbfam=\sevenmsy
  \scriptscriptfont\msbfam=\fivemsy
\def\msam#1{{\fam\msafam\relax#1}}
\def\msbm#1{{\fam\msbfam\relax#1}}


\font\tencmmib=cmmib10  \skewchar\tencmmib='177
\font\sevencmmib=cmmib7 \skewchar\sevencmmib='177
\font\fivecmmib=cmmib5 \skewchar\fivecmmib='177
\newfam\cmmibfam
\textfont\cmmibfam=\tencmmib
\scriptfont\cmmibfam=\sevencmmib
\scriptscriptfont\cmmibfam=\fivecmmib
\def\cmmib#1{{\fam\cmmibfam\relax#1}}

\def\a{\alpha}    \def\b{\beta}       
\def\c{\chi}       
\def\d{\delta}       \def\D{\Delta}    
\def\e{\varepsilon} \def\f{\phi}       
\def\F{\Phi}
\def\g{\gamma}    \def\G{\Gamma}      \def\k{\kappa}     
\def\l{\lambda}
\def\L{\Lambda}   \def\m{\mu}         \def\n{\nu}        
\def\r{\rho}
\def\vr{\varrho}  \def\o{\omega}      \def\O{\Omega}     
\def\p{\psi}
\def\P{\Psi}      \def\s{\sigma}      \def\S{\Sigma}     
\def\th{\theta}
\def\vt{\vartheta}\def\t{\tau}        \def\w{\varphi}    
\def\x{\xi}
\def\z{\zeta}
\def\CA{{\cal A}}
\def\CB{{\cal B}}
\def\CC{{\cal C}}
\def\CG{{\cal G}}
\def\CH{{\cal H}}
\def\CK{{\cal K}}
\def\CM{{\cal M}}
\def\CN{{\cal N}}
\def\CE{{\cal E}}
\def\CL{{\cal L}}\def\CJ{{\cal J}}
\def\CD{{\cal D}}
\def\CW{{\cal W}}
\def\CQ{{\cal Q}} \def\CV{{\cal V}}
\def\CP{\msbm{CP}}
\def\ep{\epsilon}
\def\C{\msbm{C}}
\def\R{\msbm{R}}
\def\BK{\cmmib{K}}

\def\rd{\partial}
\def\grad#1{\,\nabla\!_{{#1}}\,}
\def\gradd#1#2{\,\nabla\!_{{#1}}\nabla\!_{{#2}}\,}
\def\om#1#2{\omega^{#1}{}_{#2}}
\def\vev#1{\langle #1 \rangle}
\def\darr#1{\raise1.5ex\hbox{$\leftrightarrow$}
\mkern-16.5mu #1}
\def\Ha{{1\over2}}
\def\ha{{\textstyle{1\over2}}}
\def\fr#1#2{{\textstyle{#1\over#2}}}
\def\Fr#1#2{{#1\over#2}}
\def\rf#1{\fr{\rd}{\rd #1}}
\def\rF#1{\Fr{\rd}{\rd #1}}
\def\df#1{\fr{\d}{\d #1}}
\def\dF#1{\Fr{\d}{\d #1}}
\def\DDF#1#2#3{\Fr{\d^2 #1}{\d #2\d #3}}
\def\DDDF#1#2#3#4{\Fr{\d^3 #1}{\d #2\d #3\d #4}}
\def\ddF#1#2#3{\Fr{\d^n#1}{\d#2\cdots\d#3}}
\def\fs#1{#1\!\!\!/\,}   
\def\Fs#1{#1\!\!\!\!/\,} 
\def\roughly#1{\raise.3ex\hbox{$#1$\kern-.75em
\lower1ex\hbox{$\sim$}}}
\def\ato#1{{\buildrel #1\over\longrightarrow}}
\def\up#1#2{{\buildrel #1\over #2}}
\def\opname#1{\mathop{\kern0pt{\rm #1}}\nolimits}
\def\tr{\opname{Tr}}
\def\ch{\opname{ch}}
\def\Re{\opname{Re}}
\def\Im{\opname{Im}}
\def\End{\opname{End}}
\def\pr{\prime}
\def\ppr{{\prime\prime}}
\def\bs{\cmmib{s}}
\def\bbs{\bar\cmmib{s}}
\def\Dp{\rd_{\!A}}
\def\Dpp{\bar\rd_{\!A}}
\def\mapr{\!\smash{
	    \mathop{\longrightarrow}\limits^{\bs_+}}\!}
\def\mapl{\!\smash{
	    \mathop{\longleftarrow}\limits^{\bs_-}}\!}
\def\mapbr{\!\smash{
	    \mathop{\longrightarrow}\limits^{\bbs_+}}\!}
\def\mapbl{\!\smash{
	    \mathop{\longleftarrow}\limits^{\bbs_-}}\!}
\def\mapd{\Big\downarrow
 	 \rlap{$\vcenter{\hbox{$\scriptstyle \bbs_-$}}$}}
\def\mapu{\Big\uparrow
	  \rlap{$\vcenter{\hbox{$\scriptstyle\bbs_+$}}$}}
\def\ne{\nearrow}
\def\se{\searrow}
\def\nw{\nwarrow}
\def\sw{\swarrow}

\def\etal{et al.}
\def\git{/\kern-.25em/}

\def\cmp#1#2#3{Comm.\ Math.\ Phys.\ {{\bf #1}} {(#2)} {#3}}
\def\pl#1#2#3{Phys.\ Lett.\ {{\bf #1}} {(#2)} {#3}}
\def\np#1#2#3{Nucl.\ Phys.\ {{\bf #1}} {(#2)} {#3}}
\def\prd#1#2#3{Phys.\ Rev.\ {{\bf #1}} {(#2)} {#3}}
\def\prl#1#2#3{Phys.\ Rev.\ Lett.\ {{\bf #1}} {(#2)} {#3}}
\def\ijmp#1#2#3{Int.\ J.\ Mod.\ Phys.\ {{\bf #1}} 
{(#2)} {#3}}
\def\jmp#1#2#3{J.\ Math.\ Phys.\ {{\bf #1}} {(#2)} {#3}}
\def\jdg#1#2#3{J.\ Differ.\ Geom.\ {{\bf #1}} {(#2)} {#3}}
\def\pnas#1#2#3{Proc.\ Nat.\ Acad.\ Sci.\ USA.\ {{\bf #1}} 
{(#2)} {#3}}
\def\top#1#2#3{Topology {{\bf #1}} {(#2)} {#3}}
\def\zp#1#2#3{Z.\ Phys.\ {{\bf #1}} {(#2)} {#3}}
\def\prp#1#2#3{Phys.\ Rep.\ {{\bf #1}} {(#2)} {#3}}
\def\ap#1#2#3{Ann.\ Phys.\ {{\bf #1}} {(#2)} {#3}}
\def\ptrsls#1#2#3{Philos.\ Trans.\  Roy.\ Soc.\ London
{{\bf #1}} {(#2)} {#3}}
\def\prsls#1#2#3{Proc.\ Roy.\ Soc.\ London Ser.\
{{\bf #1}} {(#2)} {#3}}
\def\am#1#2#3{Ann.\ Math.\ {{\bf #1}} {(#2)} {#3}}
\def\mm#1#2#3{Manuscripta \ Math.\ {{\bf #1}} {(#2)} {#3}}
\def\ma#1#2#3{Math.\ Ann.\ {{\bf #1}} {(#2)} {#3}}
\def\cqg#1#2#3{Class. Quantum Grav.\ {{\bf #1}} {(#2)} {#3}}
\def\ivm#1#2#3{Invent.\ Math.\ {{\bf #1}} {(#2)} {#3}}
\def\plms#1#2#3{Proc.\ London Math.\ Soc.\ {{\bf #1}} 
{(#2)} {#3}}
\def\dmj#1#2#3{Duke Math.\  J.\ {{\bf #1}} {(#2)} {#3}}
\def\bams#1#2#3{Bull.\ Am.\ Math.\ Soc.\ {{\bf #1}} {(#2)} 
{#3}}
\def\jams#1#2#3{Bull.\ Am.\ Math.\ Soc.\ {{\bf #1}} {(#2)} 
{#3}}

\def\jgp#1#2#3{J.\ Geom.\ Phys.\ {{\bf #1}} {(#2)} {#3}}
\def\ihes#1#2#3{Publ.\ Math.\ I.H.E.S. \ {{\bf #1}} {(#2)} {#3}}

\def\ack{\bigbreak\bigskip\centerline{
{\bf Acknowledgements}}\nobreak}
\def\subsubsec#1{\ifnum\lastpenalty>9000\else\bigbreak\fi
  \noindent{\it #1}\par\nobreak\medskip\nobreak}

\def\submit{\baselineskip=20pt plus 2pt minus 2pt}
\def\lin#1{\medskip\noindent {${\hbox{\it #1}}$}\medskip }
\def\linn#1{\noindent $\bullet$ {\it #1} \par}


\def\V{\msbm{V}}
\def\E{\msbm{E}}
\def\CP{\msbm{CP}}
\def\R{\msbm{R}}
\def\C{\msbm{C}}
\def\Da{d_{\! A}}


\lref\TFT{
E. Witten,
{\it Topological quantum field theory},
Commun. Math. Phys. {\bf 117}  (1988) 353.
}

\lref\Wittenkaehler{
E.~Witten, 
{\it Supersymmetric Yang-Mills theory on a four-manifold},
J. Math. Phys. {\bf 35} (1994) 5101.
{\tt hep-th/9303195}.
}

\lref\GLSM{
E. Witten
{\it Phases of $N=2$ theories in two dimensions},
Nucl. Phys. {\bf B403} (1993) 159,
{\tt hep-th/9301042}.
}

\lref\Wittengwzw{
E. Witten,
{\it The $N$ matrix model and gauged WZW models},
\np{B 371}{1992}{191}.
}

\lref\Wittengr{
E.~Witten, 
{\it The Verlinde algebra and the cohomology of the Grassmannian},
{\tt hep-th/9312104}.
}

\lref\SWinv{
E.~Witten,
{\it Monopoles and four manifolds},
Math.~Research Lett.~{\bf 1} (1994) 769.
{\tt hep-th/9411102}.
}

\lref\Wittenabel{
E.~Witten,
{\it On $S$-duality in abelian gauge theory},
{\tt hep-th/9505186}.
}

\lref\MW{
G.~Moore and E.~Witten,
{\it Integration over the $u-$ plane in Donaldson theory},
{\tt hep-th/9709193}.
}

\lref\SWa{
N. Seiberg and E. Witten,
{\it Electric-magenectic duality, monopole condensation, and confinement
in $N=2$ supersymmetric Yang-Mills theory},
\np{B 426}{1994}{19}.
}

\lref\SWb{
N. Seiberg and E. Witten,
{\it Monople, duality and chiral symmetry breaking in $N=2$
supersymmetric QCD},
{\tt hep-th/9408099}.
}

\lref\VW{C. Vafa and E. Witten,
{\it A strong coupling test of S-Duality},
Nucl. Phys. {\bf B431} (1994) 3,
{\tt  hep-th/9408074}.
}


\lref\DM{
R. Dijkgraaf and G. Moore,
{\it Balanced topological field theories},
Commun. Math. Phys. {\bf 185} (1997) 411,
{\tt hep-th/9608169}.
}

\lref\DPS{
R. Dijkgraaf, J.-S. Park and B.J. Schroers,
{\it N=4 supersymmetric Yang-Mills theory on a K\"{a}hler surface}, 
{\tt hep-th/9801066}.
}

\lref\Kobayashi{
S.~Kobayashi,
{\it 
Differential geometry of complex vector bundles},
Iwanami Shoten, Publishers and Princeton Univ. Press,
1987.
}

\lref\AHS{
M.F. Atiyah, N.J. Hitchin and I.M. Singer,
{\it Self-duality in four dimensional Riemannian
geometry},
Proc.~Roy.~Soc.~London {\bf 362} (1978) {425}.
}

\lref\DK{
S.K.~Donaldson and P.B.~Kronheimer,
{\it The geometry of four-manifolds},
Clarendon Press, Oxford 1990.
}

\lref\Donaldson{
S.K.~Donaldson,
{\it Polynomial invariants for smooth $4$-manifolds},
\top{29}{1990}{257}.
}

\lref\Taubes{
C.~H.~Taubes,
{\it SW $\rightarrow$ GR. From the Seiberg-Witten equations to
pseudo-holomorphic curves},
\jams{\bf 9}{1996}{845}.
}

\lref\DonaldsonC{
S.K.~Donaldson,
{\it Anti-self-dual Yang-Mills connections on complex
algebraic surfaces and stable vector bundles},
\plms{3}{1985}{1}\semi
{\it Infinite determinants, stable bundles and curvature},
\dmj{54}{1987}{231}.
}

\lref\UY{
K.K.~Uhlenbeck and S.T.~Yau,
{\it The existence of Hermitian Yang-Mills connections on stable
bundles over K\"{a}hler manifolds}.
}

\lref\HKLR{
N.J. Hitchin, A. Karlhede, U. Lindstr\"om and M. Rocek,
{\it Hyper-K\"{a}hler metrics and supersymmetry},
\cmp{108}{1987}{535}.
}

\lref\AJ{
M. Atiyah and L. Jeffreys,
{\it Topological Lagrangians and cohomology},
\jgp{\bf 7}{1990}{119}.
}


\lref\HitchinA{
N.J. Hitchin,
{\it The self-duality equations on a Riemann surface},
\plms{3, 55}{1987}{59}.
}

\lref\HitchinB{
N.J. Hitchin,
{\it Stable bundles and integrable systems},
\dmj{54}{1987}{91}.
}

\lref\Simpson{
C.T.~Simpson,
{\it Constructing variations of Hodge structure using Yang-Mills theory
and applications to uniformization},
J.~Amer..~Math.~Soc. {\bf 1} (1988) 867.
}

\lref\SimpsonA{
C.T.~Simpson,
{\it Higgs bundles and local systems},
\ihes{75}{1992}{5}.
}

\lref\SimpsonB{
C.T.~Simpson,
{\it Moduli of representations of the fundamental group of
a smooth projective variety}, I:\ihes{79}{1994}{47};
II:\ihes{80}{1994}{5}.
}

\lref\SimpsonC{
C.T.~Simpson,
{\it The Hodge filtration on non-Abelian cohomology},
{\tt alg-geom/9604005}.
}

\lref\SimpsonD{
C.T.~Simpson,
{\it Mixed twistor structure},
{\tt alg-geom/9705006}.
}

\lref\DoMar{
R.~Donagi and E.~Markman,
{\it Spectral covers, algebraically completely integrable,
Hamiltonian systemds, and moduli of bundles},
{\tt alg-geom/9507017}.
}

\lref\Fujiki{
A.~Fujiki,
{\it HyperK\"{a}hler structure on the moduli space of flat bundles},
in "Prospects in complex geometry",
Lecture Notes in Math.~p. 1-83, Springer-Verlag, 1991.
}

\lref\Kaledin{
D.~Kaledin,
{\it Hyperk\"ahler structures on total spaces of
holomorphic cotangent bundles},
{\tt alg-geom/9710026}.
}

\lref\Hausel{
T.~Hausel,
{\it Compactification of moduli of Higgs bundles},
{\tt math.AG/9804083}.
}


\lref\Park{
J.-S.~Park,
{\it N=2 topological Yang-Mills theory
on compact K\"{a}hler surfaces},
\cmp{163}{1994}{113},
{\tt hep-th/9304060}
\semi
{\it  Holomorphic Yang-Mills theory on compact K\"{a}hler manifolds},
Nucl. Phys. {\bf B423} (1994) 559, 
{\tt hep-th/9305095}.
}

\lref\HyunPa{
S.~Hyun and J.-S.~Park,
{\it N=2 topological Yang-Mills theories
and Donaldson's polynomials},
J. Geom. Phys. {\bf 20} (1996) 31, 
{\tt hep-th/9404009}.
}

\lref\HYMDP{
S.~Hyun and J.S.~Park,
{\it Holomorphic Yang-Mills theory and the
variation of the Donaldson polynomials},
{\tt hep-th/9503092}.
}

\lref\HP{
C.~Hofman and J.-S.~Park,
{\it Sigma Models for bundles on Calabi-Yau: a proposal for
matrix string compactifications},
Nucl.~Phys.~B561 (1999) 125,
{\tt hep-th/9904150}\semi
{\it  Cohomological Yang-Mills Theories on Kahler 3-Folds},
Nucl.~Phys.~B in press, {\tt hep-th/001010}
}

\lref\ParkA{
J.-S.~ Park,
{\it Cohomological field theories on complex manifolds},
PhD thesis, 1999 Univ. of Amsterdam.
}

\lref\ParkB{
J.-S.~Park,
{\it Cohomological field theories with K\"{a}hler structure},
Adv.~Theo.~Math.~Phys.~ in press,
{\tt hep-th/9910209}
}


\lref\Yamron{
J.~Yamron, 
{\it Topological actions in twisted supersymmetric theories},
\pl{B213}{1988}{353}.
}

\lref\Wittencoft{E. Witten, 
{\it Introduction to cohomological field
theories,} 
Int. J. Mod. Phys. {\bf A6} (1991) 2775.
}

\lref\WO{
E. Witten and D. Olive, 
{\it Supersymmetry algebras that include topological charges,} 
Phys. Lett.{\bf 78B} (1978) 97.
}

\lref\LaLD{J. M. F. Labastida and C. Lozano,
{\itThe Vafa-Witten theory for gauge group SU(N)},
{\tt hep-th/9903172}.
}

\lref\LaLC{J. M. F. Labastida and C. Lozano,
{\it 
Duality in twisted N=4 supersymmetric gauge theories in four
dimensions},
Nucl.~Phys.~{\bf B537} (1999) 203,
{\tt hep-th/9806032}.
}

\lref\LaLB{
J.~M.~F.~ Labastida and C.~Lozano,
{\it 
Mass perturbations in 
twisted N=4 supersymmetric gauge
theories},
Nucl.~Phys.~ {\bf B518} (1998) 37,
{\tt hep-th/9711132}.
}

\lref\Fulton{
W.~Fulton,
{\it Intersection Theory},
Springer-Verlag 1984.
}

\lref\Monad{
J.-S.~Park,
{\it Monads and D-instantons}
Nucl.~Phys.~{\bf B 493} (1997) 198.
{\tt hep-th/9612096}.
}

\lref\HPP{
S.~Hyun, J.~Park and J.-S.~Park,
{\it Topological QCD},
Nucl.~Phys.~{\bf B 453} (1995) 119.
{\tt hep-th/9503201}
}

\lref\LMa{
J.M.F. Labastida, M. Mariño,
 Non-Abelian Monopoles on Four-Manifolds
Nucl.~ Phys.~{\bf  B448}  (1995) 373. 
{\tt hep-th/9504010}
\semi
Polynomial Invariants for SU(2) Monopoles
Nucl.~Phys.~ {\bf B456} (1995) 633.
{\tt hep-th/9507140}
}

\lref\MMa{
Marcos Marino, Gregory Moore,
Integrating over the Coulomb branch in N=2 gauge theory,
Nucl.~Phys.~Proc.~Suppl.~ {\bf 68} (1998) 336.
{\tt hep-th/9712062}
}

\lref\MMb{
Marcos Marino, Gregory Moore,
{\it 
The Donaldson-Witten function for gauge groups of rank larger than one},
Commun.~Math.~Phys.~{\bf  199} (1998) 25.
{\tt hep-th/9802185},
}

\lref\MMc{
Marcos Marino, Gregory Moore,
Donaldson invariants for nonsimply connected manifolds
{\tt hep-th/9804104}
}

\lref\Wang{
S.~Wang, {\it A vanishing theorem for Seiberg-Witten invariants},
Math.~Res.~Letters {\bf 2} (1995) 305.
}

\lref\FS{
R.~Fintushel and R.J.~Stern,
{\it Nondiffeomophic symplectic 4-manifolds with the
same Seiberg-Witten invariants,
}
{\tt math.SG/9811019}.
}

\lref\SYZ{
A. Strominger, S.-T. Yau and E. Zaslow,
{\it Miror symmetry is a $T$-duality},
\np{B 479}{1996}{243},
{\tt hep-th/9612121}.
}

\lref\Vafa{
C. Vafa, {\it Extending mirror conjecture to Calabi-Yau with bundles},
{\tt hep-th/9804131}.
}

\lref\CeVa{
S. Cecotti and C. Vafa, 
{\it Topological anti-topological fusion,} 
Nucl. Phys. {\bf B367} (1991) 359.
}

\lref\LMS{ 
A.~Losev, N.~Nekrasov and S.~Shatashvili,
{\it Issues in topological gauge theory},
Nucl.~Phys.~{\bf B534} (1998) 549.
{\tt hep-th/9711108} 
}

\newsec{Introduction}

It is more than a decade ago that  Witten introduced
a quantum field theoretic formulation \TFT\ of the  four-dimensional 
differential topological invariants of 
Donaldson \Donaldson\DK.  
In this approach Donaldson invariants are defined as
certain correlation functions of twisted $N=2$ spacetime 
supersymmetric Yang-Mills (SYM) theory in four dimensions.
This approach
eventually opened up a new horizon in mathematics via the
quantum properties of underlying physical theory,  which is
uncovered by
Seiberg and Witten \SWa\SWb.  The resulting Seiberg-Witten
invariants are much more simple, while carrying the
same information for the smooth structure of four manifolds
as the Donaldson-Witten invariants \SWinv. The Donaldson-Witten
theory can be generalized by twisting more general $N=2$
SYM theory with hypermultiplets \HPP\LMa. Such a theory can be also
solved by the physical solution of the underlying $N=2$
SYM theory \MW\LMS\MMa\MMb\MMc. 
However, those invariants defined by such
a theory carry the same information as Seiberg-Witten invariants
for the smooth structure of four manifolds.
It is also shown that
Gromov-Witten invariants are equivalent to Seiberg-Witten
invariants \Taubes.

On the other hand, it has been demonstrated that the Seiberg-Witten invariant
alone does not distinguish smooth structures of certain, 
at least non-simply-connected, four manifolds belong to 
a same homeomorphism class \Wang\FS.
Thus it is a challenging open problem to define four manifold
invariants beyond Donaldson-Witten or Seiberg-Witten
invariants. 

The purpose of this paper is to propose 
candidates of such invariants restricting to the K\"{a}hler
cases. We will take quantum field theoretic approach
to these invariants. Our generalization of   Donaldson-Witten
theory involves the moduli space of stable Higgs bundles of
Simpson \Simpson\SimpsonA\SimpsonB\ instead of
the moduli space of stable bundles.
We also propose the similar generalization of  Vafa-Witten theory \VW\
- a twisted $N=4$
SYM theory \Yamron, as well as related generalization of 
Seiberg-Witten theory. We conjecture that the equivalence
between Donaldson-Witten and Seiberg-Witten invariants
remains valid to their generalized versions. 
We follow a general approach to constructing 
cohomological field theory with a K\"{a}hler structure,
developed in \ParkB.
Compared with Donaldson-Witten or Vafa-Witten theories
our models do not have underlying spacetime supersymmetric
theory.
Our models, nevertheless, are "connected" to 
the physical $N=2$ or $N=4$
SYM theories by certain renormalization group flows,
and those physical theories reside  in particular of fixed points.
Our conjecture is that the other fixed points give rise to
new invariants of four manifold beyond Donaldson-Witten
and Seiberg-Witten invariants.

\newsec{Generalized Donaldson-Witten Theory}

Donaldson-Witten theory (twisted $N=2$ SYM) on  a K\"{a}hler surface   
\Park\Wittenkaehler\HyunPa\ 
is an example of $N_c=(2,0)$ supersymmetric
gauged sigma model in zero-dimensions \ParkB.
Such a model is classified by a K\"{a}hler target space
$\CA$ with a group $\CG$ acting as an isometry, which determines
a $\CG$-equviariant momentum map $\m:\CA \rightarrow Lie(\CG)^*$.
We further have a  Hermitian holomorphic vector bundle $\msbm{E}\rightarrow 
\CA$ with $\CG$-equivariant holomorphic 
section $\eufm{S}$. Then the bosonic part of the path integral reduces 
to, provided that we
are evaluating correlation functions for supersymmetric observables,
an integration over  $\CM := \eufm{S}^{-1}(0)\cap \m^{-1}(\zeta)/\CG$;
the solution space of the following
equations, modulo $\CG$,
\eqn\lcaa{
\eqalign{
\eufm{S}=0,\cr
\m -\zeta=0,\cr
}
}
where $\zeta$ is the Fayet-Iliopoulos (FI) term. 
Those observables correspond to elements of $\CG$-equivariant
cohomology of $\CA$.
The correlation functions of such observables
are identified with intersection numbers of homology
cycles, represented by the observables, in $[e(\V)]$, 
where $[e(\V)]$ denotes the cycle in $\CM$ Poincar\'e dual 
to the Euler class $e(\V)$ of 
the anti-ghost bundle $\V$ over $\CM$. 
If the model has actually $N_c=(2,2)$
supersymmetry the anti-ghost bundle $\V$ can be identified with
the tangent bundle $T\CM$ and the partition function is the
Euler characteristic of $\CM$.
The moral  underlying cohomological
field theory is  that the triple $(\CA,\CG,\msbm{E})$
can be all infinite dimensional but certain path integral
can still be reduced to an integral over finite dimensional space
$\CM$.

In Donaldson-Witten theory
$\CA$  is the space of all  connections (gauge fields) on a Hermitian vector 
bundle $E\rightarrow M$ over a
complex $2$-dimensional K\"{a}hler manifold $M$ with K\"{a}hler
form $\o$ and
$\CG$ is the group of all gauge transformations.
This determines a localization equation from the momentum
map $\m$;
\eqn\lcb{
i F\wedge \o -\Fr{\zeta}{2}\o^2 I_E =0,
}
The solution
space of this equation modulo $\CG$ is infinite dimensional. 
Thus we consider an infinite dimensional  
bundle $\msbm{E}\rightarrow \CA$ with $\CG$-equivariant holomorphic 
section $\eufm{S}$. We introduce a
complex structure of $\CA$
by declaring that the $A^{0,1}$ component of a connection
$1$-form $A =A^{1,0} + A^{0,1}$ 
represents holomorphic coordinates.
Then there is an unique choice
$\eufm{S}= F^{0,2}$ on a general K\"{a}hler manifold, leading
to another localization equation,
\eqn\lcbb{
F^{0,2}=0.
}
An integrable connection $F^{0,2}=\Dpp^2=0$ 
is called Einstein-Hermitian or Hermitian-Yang-Mills 
if it further satisfies \lcb. 
Thus the path integral is localized to the moduli space $\CM_{EH}$
of Hermitian-Yang-Mills connections, or equivalently, 
a result due to \DonaldsonC\UY,
the moduli space of semi-stable holomorphic bundles on
$M$. 
In this case the anti-ghost bundle $\V \rightarrow \CM$
is trivial, due to Donaldson, 
and the correlation functions of supersymmetric
observables can be interpreted as certain intersection
pairings of homology cycles in the moduli space $\CM_{EH}$.

In the section we generalize Donaldson-Witten theory on K\"{a}hler surfaces.
The basic idea
is to extend our target space $\CA$ of the $N_c=(2,0)$
model - corresponding to Donaldson-Witten theory,
to the total space $T^*\CA$ of the cotangent bundle of $\CA$.
Since $\CA$ is a flat affine K\"{a}hler manifold
a cotangent vector is represented as an element of
$\O^1(M, End(E))$.
Thus we introduce additional bosonic fields $\w$ given
by an adjoint valued $1$-form $\w\in \O^1(M,End(E))$.
Then, by decomposing 
 $\w =\w^{1,0} +\w^{0,1}$, we have to  declare $\w^{1,0}$ 
to represent
holomorphic coordinates on the fiber space of $T^*\!\CA$,
since we already fixed a complex structure of $\CA$
by declaring that the $A^{0,1}$ component of a connection
$1$-form represents holomorphic coordinates.
A beautiful fact
for any cotangent bundle of a K\"{a}hler manifold
is that it always has canonical hyper-K\"{a}hler structure \Kaledin.
Thus  it is natural to consider hyper-K\"{a}hler quotients
of $T^*\!\CA$ by $\CG$;
\eqn\lcbd{
\eqalign{
\Dp^* \w^{1,0}=0,\cr
i\L\left( F + [\w^{1,0},\w^{0,1}]\right) - \zeta I=0.\cr
}
}
However
the resulting hyper-K\"{a}hler quotient space is
infinite dimensional.\foot{In one complex dimensions 
the hyper-K\"{a}hler quotient space is Hitchin's
moduli space \HitchinA.}
To obtain a
finite dimensional space we extend the bundle $\msbm{E}\rightarrow
\CA$ to $\widetilde \msbm{E}\rightarrow T^*\CA$ and try
to cut out the hyper-K\"{a}hler quotient space by the
vanishing locus of suitable $\CG$-equivariant
holomorphic sections. 
A natural choice on a K\"{a}hler surface
is
\eqn\lcbe{
\eqalign{
F^{0,2}=0,\cr
\Dpp \w^{1,0}=0,\cr
\w^{1,0}\wedge \w^{1,0}=0,\cr
}
}
which defines Higgs bundles of Simpson \Simpson\SimpsonA.
The above equation can be viewed as a generalization
of the integrability $\Dpp^2=0$ of the connection $\Dpp$
to the integrability of the extended connection 
$\cmmib{D}^\ppr =\Dpp +\w^{1,0}$.
Our model based on \lcbe\ is a 
generalization of  Donaldson-Witten theory.

Another beautiful fact
for any cotangent bundle of a K\"{a}hler manifold
is that it always has the equivariant $S^1$-action
acting on the fiber. Such a $S^1$-action on $T^*\! \CA$
descends to the  moduli spaces above.
We will use the $S^1$ symmetry to define
a family of models, which have many interesting
limits.

\subsec{Preliminaries}

We consider a rank $r$ Hermitian vector bundle $E\rightarrow M$
over a complex $d$-dimensional K\"{a}hler manifold $M$ with 
K\"{a}hler form $\o$. Consider the space $\CA$ 
of all connections of $E$
and the cotangent bundle $T^*\CA$.
First we determine the fields representing 
the cotangent space $T^*\CA$.
For the base space $\CA$ of $T^*\!\CA$ 
we have connection $1$-form $A = A^{1,0}+A^{0,1}$
with the usual gauge transformation law.
We introduce a complex structure $I$ on $\CA$
using the complex structure of $M$ by declaring $A^{0,1}$
to represent holomorphic coordinates.
Since $\CA$ is a flat affine K\"{a}hler manifold
a cotangent vector is represented as an element of
$\O^1(M, End(E))$.
We introduce
an adjoint valued bosonic $1$-form $\w\in \O^1(M,End(E))$,
which may  be regarded as an element of the cotangent space of $\CA$.
According to the complex structure of $M$ we have
a decomposition $\w =\w^{1,0}+\w^{0,1}$.
Then it is natural to fix the 
complex structure of the fiber space of $T^*\!\CA$ 
by declaring $\w^{1,0}$
to be a holomorphic coordinate. Thus the (holomorphic)
tangent space of $T^*\CA$ is
given by
\eqn\lcbb{
\O^{0,1}(M, End(E))\oplus \O^{1,0}(M, End(E)).
}
We denote the above complex structure also by  $I$ and call
it the
preferred complex structure, which has been induced from the
complex structure of $M$.
The total   K\"{a}hler potential   $\CK(A,\w)$ of the 
total space $T^*\CA$ is
given by
\eqn\lcd{
 \CK(A,\w) = \CK(A) -
\Fr{i}{2 (d)!\pi^2}\int_M \tr(\w^{1,0}\wedge \w^{0,1})\wedge\o^{d-1}
}
where the K\"aher potential $\CK(A)$ of $\CA$ is
\eqn\lce{
\CK(A) = 
\Fr{1}{4 (d)!\pi^2}\int_M \k \tr (F\wedge F)\wedge\o^{d-2}.
} 
and the added term is a K\"{a}hler potential in the space
$\CB$.  On the total space $T^*\CA$ we have a obvious action
of the infinite dimensional group $\CG$ of all gauge transformations,
preserving the K\"{a}hler potential $\CK(A,\w)$.

Now we introduce our $N_c=(2,0)$ supercharges $\bs_+$ and
$\bbs_-$ with the familiar commutation relations
\eqn\lcf{
\bs_+^2=0,\qquad \{\bs_+,\bbs_+\} = -i \phi^a_{++}\CL_a,
\qquad \bbs_+^2=0.
}
The supercharges are identified with the differentials of
$\CG$-equivariant cohomology of our target space $T^*\CA$.
Thus $\phi^a_{++}\CL_a$ is the infinitesimal gauge transformation
generated by the adjoint scalar $\phi_{++} \in Lie(\CG) = \O^0(M,End(E))$.
{}From the complex structure of $T^*\CA$ introduced above
we have two sets of  holomorphic multiplets 
$(A^{0,1}, \p^{0,1}_+)$ and $(\w^{1,0},\l^{1,0}_+)$
and their anti-holomorphic partners.
The supersymmetry transformation laws are given by
\eqn\lcg{
\eqalign{
\bs_+ A^{0,1} &=i\p^{0,1}_+,\cr
\bbs_+ A^{0,1}&=0,\cr
\bs_+ A^{1,0}&=0,\cr
\bbs_+ A^{1,0}&=i\bar\p^{1,0}_+,\cr
}\qquad
\eqalign{
\bs_+ \p^{0,1}_+&=0,\cr
\bbs_+\p^{0,1}_+ &=- \Dpp \phi_{++},\cr
\bs_+\bar\p^{1,0}_+&=-\Dp \phi_{++},\cr
\bbs_+\bar\p^{1,0}_+&=0,\cr
}
}     
and
\eqn\lch{
\eqalign{
\bs_+ \w^{1,0}&=i\l^{1,0}_+,\cr
\bbs_+ \w^{1,0}&=0,\cr
\bs_+ \w^{0,1} &=0,\cr
\bbs_+ \w^{0,1}&=i\bar\l^{0,1}_+,\cr
}\qquad
\eqalign{
\bs_+\l^{1,0}_+&=0,\cr
\bbs_+\l^{1,0}_+&=[\phi_{++}, \w^{1,0}],\cr
\bs_+ \bar\l^{0,1}_+&=[\phi_{++}, \w^{0,1}],\cr
\bbs_+\bar\l^{0,1}_+ &= 0.\cr
}
}

{}From the transformation laws we have the
following total $\CG$-equivariant K\"{a}hler form
on $T^*\!\CA$,
\eqn\lci{
\eqalign{
\widehat\varpi^\CG_T=&
i\bs_+\bbs_+ \CK(A,\w)\cr
=& \Fr{i}{2(d)!\pi^2}\int_M \tr\left(\phi_{++} 
\left(F + [\w^{1,0},\w^{0,1}]\right)
\right)
\wedge \o^{d-1} 
\cr
&
+\Fr{1}{2(d)!\pi^2}
\int_M \tr\left(\p^{0,1}_+\wedge\bar\p^{1,0}_+
+\l^{1,0}_+\wedge\bar\l^{0,1}_+
\right)\wedge \o^{d-1}.
}
}
The second term in the above is the K\"{a}hler form $\varpi$
and the first term is the real $\CG$-momentum map $\phi^a_{++}\m_a$,
$\m_\msbm{R}:T^*\CA\rightarrow Lie(\CG)^*=\O^{2n}(M,End(E))$;
\eqn\lcj{
\eqalign{
\m_\msbm{R} 
&=  \Fr{1}{2(d)!\pi^2} \left(F +[\w^{1,0},\w^{0,1}]\right)\wedge \o^{d-1}\cr
&= \Fr{1}{2d (d)!\pi^2}\L\left( F + [\w^{1,0},\w^{0,1}]\right) \o^d,
}
}
where $\L$ denote the adjoint of wedge multiplication with $\o$.

Following Hitchin \HitchinA\  we have a natural
hyper-K\"{a}hler structure $I$, $J$ and $K$ on
$T^*\!\CA$. Note that the additional complex structures
$J$ and $K$ have no relation with the complex
structure on the manifold $M$. 
Then we define the holomorphic symplectic
form $\varpi_\msbm{C}$ on $T^*\CA$ by
\eqn\lck{
\eqalign{
\varpi_\msbm{C}((\d_1 A^{0,1},&\d_1 \w^{1,0}), 
(\d_2 A^{0,1},\d_2 \w^{1,0}))
\cr
&
= \Fr{1}{2(d)!\pi^2}
\int_M\tr\left(\d_2 \w^{1,0}\wedge *\d_1 A^{0,1} - \d_1 \w^{1,0}\wedge
*\d_2 A^{0,1}\right).
}
}
The corresponding complex momentum map $\m_\msbm{C}$ on
$T^*\!\CA$ is
given by
\eqn\lcm{
\m_\msbm{C} = \Fr{1}{2(d)!\pi^2}\Dpp \w^{1,0}\wedge\o^{d-1}
=\Fr{1}{2d\cdot (d)!\pi^2}(\L\Dpp \w^{1,0})\wedge\o^d.
} 
Using the K\"{a}hler identities
\eqn\kaehler{
\Dpp^* =i[\Dp,\L],\qquad
\Dp^*=-i[\Dpp,\L],
}
we see that the zeros of the complex momentum map is given by
\eqn\lcn{
\L\Dpp \w^{1,0} \rightarrow\Dp^* \w^{1,0}=0.
}

We again consider a rank $r$ Hermitian vector bundle $E\rightarrow M$
over a complex $2$-dimensional K\"{a}hler manifold $M$ with 
K\"{a}hler form $\o$. Consider the space $\CA$ 
of all connections on $E$
and the cotangent bundle $T^*\CA$.
We have the same holomorphic coordinates fields $A^{0,1}$
and $\w^{1,0}\in \O^{1,0}(M,End(E))$ of $T^*\CA$, with the supersymmetry
transformation laws in \lcg\ and \lch. 
We also have the usual $N_c=(2,0)$ gauge multiplet.

\subsec{Our Model}

Now consider
an infinite dimensional $\CG$-equivariant 
holomorphic Hermitian vector
bundle $\widetilde\msbm{E}\rightarrow T^*\!\CA$ over $T^*\!\CA$
with a suitable $\CG$-equivariant holomorphic section 
$\widetilde\eufm{S}(A^{0,1},\w^{1,0})$,
i.e., $\bbs_+ \widetilde\eufm{S}(A^{0,1},\w^{1,0})=0$.
We only have the following possibility for this;
\eqn\slxa{
\widetilde\eufm{S}(A^{0,1},\w^{1,0}) = F^{0,2}\oplus 
\Dpp \w^{1,0}\oplus
(\w^{1,0}\wedge \w^{1,0}).
}
We choose this most general form as
our holomorphic section.  
We have a natural
paring of  the holomorphic section
with corresponding anti-ghost fields $\Upsilon_-$
given by $\int_M\tr (\Upsilon_- \wedge *  \eufm{S})$.
Thus the anti-ghost for the $F^{0,2}$ bit of section
belongs to $\O^{2,0}(M,End(E))$, the anti-ghost
for the mixed part belongs to $\O^{1,1}(M,End(E))$ and the anti-ghost
for $(\w^{1,0}\wedge \w^{1,0})$ belongs to $\O^{0,2}(M, End(E))$.
Associated with the holomorphic section $F^{0,2}$ over
the base space $\CA$ of $T^*\!\CA$ we have
Fermi multiplet $(\chi^{2,0}_-, H^{2,0})\in \O^{2,0}(M,End(E))$
and anti-Fermi multiplet $(\bar\chi^{0,2}_-, H^{0,2})$,
\eqn\slxc{
\eqalign{
\bs_+\chi^{2,0}_- &= -H^{2,0},\cr
\bbs_+\chi^{2,0}_-&= 0,\cr
\bs_+\bar\chi^{0,2}_- &=0,\cr
\bbs_+\chi^{0,2}_-&= -H^{0,2},\cr
}
\qquad
\eqalign{
\bs_+H^{2,0} &=0,\cr
\bbs_+ H^{2,0} &=  -i[\phi_{++},\chi^{2,0}_-]
        , \cr
\bs_+ H^{0,2} &=  -i[\phi_{++},\bar\chi^{0,2}_-]
        , \cr
\bbs_+H^{0,2} &=0.\cr
}
}
Associated with the mixed component of holomorphic section $\Dpp \w^{1,0}$ 
over $\CB$ of $T^*\!\CA$ we have
Fermi multiplets $(\chi^{1,1}_-, H^{1,1})\in \O^{1,1}(M,End(E))$
and their anti-Fermi partners $(\bar\chi^{1,1}_-, \bar H^{1,1})$,
\eqn\slcp{
\eqalign{
\bs_+\chi^{1,1}_- &= -H^{1,1},\cr 
\bbs_+\chi^{1,1}_-&=0,\cr
\bs_+\bar\chi^{1,1}_-&=0,\cr
\bbs_+\bar\chi^{1,1}_-&= -\bar H^{1,1},\cr
}\qquad
\eqalign{
\bs_+ H^{1,1}&=0,\cr
\bbs_+ H^{1,1}&=-i[\phi_{++},\chi^{1,1}_-],\cr
\bs_+ \bar H^{1,1}&=-i[\phi_{++},\bar\chi^{1,1}_-],\cr
\bbs_+ \bar H^{1,1}&=0.
}
}
Associated with the holomorphic section $\w^{1,0}\wedge \w^{1,0}$ 
over the fiber space $\CB$ of $T^*\!\CA$ we have
Fermi multiplet $(\eta^{0,2}_-, K^{0,2})\in \O^{2,0}(M,End(E))$
and their anti-Fermi partner $(\bar\eta^{2,0}, H^{2,0})$
\eqn\slxd{
\eqalign{
\bs_+\eta^{0,2}_- &= -K^{2,0},\cr
\bbs_+\eta^{0,2}_-&= 0,\cr
\bs_+\bar\eta^{2,0}_- &= 0,\cr
\bbs_+\bar\eta^{2,0}_-&= -K^{0,2},\cr
}
\qquad
\eqalign{
\bs_+K^{2,0} &=0,\cr
\bbs_+ K^{2,0} &=  -i[\phi_{++},\eta^{0,2}_-]
        , \cr
\bbs_+ K^{0,2} &=  -i[\phi_{++},\bar\eta^{2,0}_-]
        , \cr
\bs_+K^{0,2} &=0.\cr
}
}

Now we consider the following $N_c=(2,0)$
supersymmetric action functional 
\eqn\slcpp{
\eqalign{
S = &
\Fr{\bs_+\bbs_+}{4\pi^2}\int_{M}\!\tr\biggl(
\phi_{--} \left(F \wedge\o+ [\w^{1,0},\w^{0,1}]\wedge\o 
+\Fr{i\zeta}{2}\o^2 I_E \right) 
+\eta_- * \bar\eta_-\biggr)
\cr
&
+\Fr{\bs_+\bbs_+}{4\pi^2} \int_{M}\!\tr\left(
\c^{2,0}_-\!\wedge *\bar\c^{0,2}_-
+\c^{1,1}_-\wedge *\bar\c^{1,1}_-
+\eta^{0,2}_-\!\wedge *\bar\eta^{2,0}_-
\right)
\cr
&
+\Fr{i\bs_+}{4\pi^2}\int_{M} \tr \biggr(
\c^{2,0}_-\wedge *F^{0,2} 
+\c^{1,1}_-\wedge *  \Dpp \w^{1,0}
+\eta^{0,2}_-\wedge *  (\w^{1,0}\wedge \w^{1,0})\biggr)
\cr
&
+\Fr{i\bbs_+}{4\pi^2}\int_M\tr\biggl(
\bar\c^{0,2}_-\wedge * F^{2,0}
+\bar\c^{1,1}_-\wedge * \Dp \w^{0,1}
+\bar\eta^{2,0}_-\wedge * (\w^{0,1}\wedge \w^{0,1})\biggr),
}
}
We set $\zeta=0$ for simplicity by restricting to the
case with $c_1(E)=0$.
By expanding the above and integrating out
the auxiliary fields we see that the path integral is
localized to the moduli space defined by the following
equations
\eqn\noaga{
\eqalign{
F^{0,2}=0,\cr
\w^{1,0}\wedge \w^{1,0}=0,\cr
\Dpp \w^{1,0}=0,\cr
i\L (F +[\w^{1,0},\w^{0,1}]) -\zeta I_E=0.
}
}
The first three equations above are from $\tilde\eufm{S}=0$
and the last equation is from the total momentum map $\m_\msbm{R}$
\lcj.
The   Higgs bundle $(\Dpp,\w^{1,0})$ of Simpson \Simpson\SimpsonA\
is defined by the first three equations in \noaga, i.e. the equations
in \lcbe.
Those  equations can be regarded as
integrability $(\cmmib{D}^\ppr)^2=0$ of the  extended half "connection"
$\cmmib{D}^\ppr=\Dpp + \w^{1,0}$. 
There is notion of semi-stable Higgs bundle and
a theorem analogous to Donaldson-Uhlenbeck-Yau such
that every semi-stable 
Higgs bundle $(E, \w^{1,0})$ has an  Einstein-Hermitian
metric;
\eqn\nob{
i\L(F + [\w^{1,0},\w^{0,1}]) - \zeta I_{E}=0.
}
Furthermore the extended connection is flat 
$\cmmib{D}^\pr\circ \cmmib{D}^\ppr+\cmmib{D}^\ppr\circ  
\cmmib{D}^\pr=0$ if and only if
$c_1(E,\w^{1,0}) = c_2(E,\w^{1,0})=0$.
Thus the path integral of our model is localized
to the moduli space of semi-stable Higgs bundles.
We also have other bosonic localization equations, as usual
\eqn\asusau{
\eqalign{
\Da \phi_{++}=0,\cr
[\phi_{++}, B]=0,\cr
[\phi_{++},\phi_{--}]=0.\cr
}
}
If the connections are irreducible we have $\phi_{\pm\pm}=0$
and $\CG$ acts freely on the solution space of \noaga.
The resulting moduli space is then isomorphic to the
moduli space of stable Higgs bundles.
We denote the moduli space of semi-stable
Higgs bundle by $\CN$. Note that the moduli space
$\CN$ contains the moduli space $\CM$ of semi-stable
bundles, equivalently the moduli space of EH or
anti-self-dual connections on a K\"{a}hler surface $M$.

{}From now on we set $\zeta=0$ for simplicity.

\subsec{Comparison with Donaldson-Witten Theory}

At this point it is useful to compare with 
Donaldson-Witten theory. The path integral
of Donaldson-Witten theory is localized to the
moduli space $\CM$ of anti-self-dual  connections
defined by
\eqn\tola{
\eqalign{
(\Dpp)^2=0,\cr
\L(\Dp\circ\Dpp + \Dpp\circ\Dp) =0.\cr
}
}
Define $\cmmib{D}^\ppr = \Dpp+ \w^{1,0}$ and 
$\cmmib{D}^\pr = \Dp + \w^{0,1}$. 
Our localization equations \noaga\ can be
written as
\eqn\tolb{
\eqalign{
(\cmmib{D}^\ppr)^2=0,\cr
\L(\cmmib{D}^\pr\circ \cmmib{D}^\ppr 
+ \cmmib{D}^\ppr\circ \cmmib{D}^\pr )=0.\cr
}
}
Similarly we can combine the superpartners of $A^{0,1}$ and
$\w^{1,0}$, and the anti-ghosts $(\chi^{2,0},\chi^{1,1}_-,\eta^{0,2}_-)$.
To see this let us define extended fields
\eqn\juicy{
\eqalign{
\cmmib{A}^{(0,1)}&:= A^{0,1}+ \w^{1,0},\cr
\Psi^{(0,1)}_+ &:= \p^{0,1}_+ + \l^{1,0}_+,\cr
\Upsilon_-^{(2,0)} &:= \chi^{2,0}_- + \chi^{1,1}_- + \eta^{0,2}_-,\cr
\cmmib{H}^{(2,0)} &:= H^{2,0} +  H^{1,1} + K^{0,2},\cr
}\qquad
\eqalign{
\cmmib{A}^{(1,0)} &:= A^{1,0}+ \w^{0,1},\cr
\bar\Psi^{(1,0)}_+ &:= \bar\p^{1,0}_+ + \bar\l^{0,1}_+,\cr
\bar\Upsilon_-^{(0,2)} &:= \bar\chi^{0,2}_- + \bar\chi^{1,1}_- 
+ \bar\eta^{2,0}_-,\cr
\cmmib{H}^{(0,2)} &:= H^{0,2} + \bar H^{1,1} + K^{2,0},\cr
}
}
where the superscript of the extended fields represent
a graded form degree on $M$. That is we exchange holomorphic
and antiholomorphic differential form degree on $M$ of fields associated
with $\w^{1,0}$ and $\w^{0,1}$. For example
the extended anti-ghost  $\Upsilon^{(2,0)}_-$
is associated with the total holomorphic section
$\widetilde\eufm{S} :=\cmmib{F}^{(0,2)}:= F^{0,2} + \Dpp \w^{1,0} 
+ \w^{1,0}\wedge \w^{1,0}$  of $\widetilde \msbm{E}\rightarrow T^*\!\CA$
by the pairing 
$\int_M \tr \left(\Upsilon^{(2,0)}_-\wedge * \cmmib{F}^{(0,2)}\right)$. 
Note that the combinations \juicy\
preserve the ghost numbers
\eqn\juigh{
\eqalign{
\Psi^{(0,1)}_+ &:(+1,0),\cr
\Upsilon_-^{(2,0)} &:(-1,0),\cr
}\qquad
\eqalign{
\bar\Psi^{(1,0)}_+ &:(0,+1),\cr
\bar\Upsilon_-^{(0,2)} &:(0,-1) .
}
}
The supersymmetry transformation laws for
the coordinate fields of $T^*\!\CA$ are,
combining \lcg\ and \lch,
\eqn\juia{
\eqalign{
\bs_+ \cmmib{A}^{(0,1)} &=i\P^{(0,1)}_+,\cr
\bbs_+ \cmmib{A}^{(0,1)}&=0,\cr
\bs_+ \cmmib{A}^{(1,0)}&=0,\cr
\bbs_+ \cmmib{A}^{(1,0)}&=i\bar\P^{(1,0)}_+,\cr
}\qquad
\eqalign{
\bs_+ \P^{(0,1)}_+&=0,\cr
\bbs_+\P^{(0,1)}_+ &=- \cmmib{D}^\ppr \phi_{++},\cr
\bs_+\bar\P^{(1,0)}_+&=-\cmmib{D}^\pr \phi_{++},\cr
\bbs_+\bar\P^{(1,0)}_+&=0.\cr
}
}     
The supersymmetry transformation laws for the Fermi
multiplet $(\Upsilon^{(2,0)}_-, \cmmib{H}^{(2,0)})$
are, by combining \slxc, \slcp, and \slxd\ together,
\eqn\juib{
\eqalign{
\bs_+\Upsilon^{(2,0)}_- &= -\cmmib{H}^{(2,0)},\cr 
\bbs_+\Upsilon^{(2,0)}_-&=0,\cr
\bs_+\bar\Upsilon^{(0,2)}_-&=0,\cr
\bbs_+\bar\Upsilon^{(0,2)}_-&= -\cmmib{H}^{(0,2)},\cr
}\qquad
\eqalign{
\bs_+ \cmmib{H}^{(2,0)}&=0,\cr
\bbs_+ \cmmib{H}^{(2,0)}&=-i[\phi_{++},\Upsilon^{(2,0)}_-],\cr
\bs_+ \cmmib{H}^{(0,2)}&=-i[\phi_{++},\bar\Upsilon^{(0,2)}_-],\cr
\bbs_+\cmmib{H}^{(0,2)}&=0.
}
}
We have the usual $N_c=(2,0)$ gauge multiplet 
associated with the unitary gauge transformation.
For convenience we rewrite down 
supersymmetry the transformation laws
\eqn\juic{
\eqalign{
\bs_+ \phi_{--} = i\eta_-,\cr
\bbs_+\phi_{--}=i\bar\eta_-,\cr
}
\qquad
\eqalign{
\bs_+ \eta_-&=0,\cr
\bbs_+\eta_- &=+i H_0 + \Fr{1}{2}[\phi_{++},\phi_{--}],\cr
\bs_+\bar\eta_-&=-i H_0 +\Fr{1}{2}[\phi_{++},\phi_{--}],\cr
\bbs_+\bar\eta_-&=0,
}\qquad \eqalign{
\bs_+\phi_{++}=0,\cr
\bbs_+\phi_{++}=0.\cr
}
}

Now the action functional $S$ in \slcpp\ can be
written as
\eqn\juid{
\eqalign{
S = &
\Fr{\bs_+\bbs_+}{4\pi^2}\int_{M}\!\tr\biggl(
\phi_{--}\cmmib{F}\biggr)\wedge\o 
\cr
&
+\Fr{\bs_+\bbs_+}{4\pi^2} \int_{M}\!\tr\left(
\eta_-\wedge * \bar\eta_- 
+\Upsilon^{(2,0)}_-\!\wedge *\bar\Upsilon^{(0,2)}_-\right)
\cr
&
+\Fr{i\bs_+}{4\pi^2}\int_{M} \tr \biggr(
\Upsilon^{(2,0)}_-\wedge * \cmmib{F}^{(0,2)} 
\biggr)
+\Fr{i\bbs_+}{4\pi^2}\int_M\tr\biggl(
\bar\Upsilon^{(0,2)}_-\wedge *\cmmib{F}^{(2,0)}\biggr),
}
}
where
\eqn\gfoit{
\m_\msbm{R} = \Fr{1}{4\pi^2}\cmmib{F}\wedge\o.
}
The above action functional has exactly same form as 
Donaldson-Witten theory on K\"{a}hler $2$-folds.
We remark that the K\"{a}hler identities \kaehler\ 
are important technical tools in analyzing Donaldson-Witten
theory on K\"{a}hler manifolds.
Simpson showed that one also has
the K\"{a}hler identities for Higgs bundles,
\eqn\juie{
(\cmmib{D}^\pr)^* = i[\L, \cmmib{D}^\ppr],
\qquad (\cmmib{D}^\ppr)^* = -i[\L ,\cmmib{D}^\pr],
}
We will work with the above shorthand notations.

\subsec{The Path Integral and New Invariants of Four-Manifolds}

The explicit form of the total action functional $S^\pr$
after integrating out all the auxiliary fields from $S$ 
is given by
\eqn\july{
\eqalign{
S^\pr=&\Fr{1}{4\pi^2}\int_M \tr\biggl(
-\Fr{1}{2} \cmmib{F}^+ \wedge * \cmmib{F}^+
-  \cmmib{D}\phi_{++}* \cmmib{D}\phi_{--}
+\Fr{1}{4}[\phi_{++},\phi_{--}]*[\phi_{++},\phi_{--}]
\cr
&
+\phi_{--}*\L[\P^{(0,1)}_+,\bar\P^{(1,0)}_+]
+i\Upsilon^{2,0}_-\wedge *[\phi_{++},\bar\Upsilon^{(0,2)}_-]
+i[\phi_{++},\eta_-]*\bar\eta_-
\phantom{\biggr)}
\cr
&
-i \cmmib{D}^\pr\bar\eta_- \wedge *  \P^{(0,1)}_+
-i \cmmib{D}^\ppr\eta_-\wedge * \bar\P^{(1,0)}_+
-\Upsilon^{2,0}\wedge * \cmmib{D}^\ppr\P^{(0,1)}_+
\phantom{\biggr)}
\cr
&
-\bar\Upsilon^{(0,2)}\wedge  * \cmmib{D}^\pr \bar\P^{(1,0)}_+
\biggr),
}
}
where $\cmmib{D}=\cmmib{D}^\pr +\cmmib{D}^\ppr$
and we we used the extended K\"{a}hler identities \juie.
We also used notation $\cmmib{F}^+$, which is given by
\eqn\juiccd{
\cmmib{F}^+ =\cmmib{F}^{(2,0)} + \Fr{1}{2}(\L \cmmib{F})\o 
+ \cmmib{F}^{(0,2)},
}
so that $\cmmib{F}^{+}|_{\w^{1,0}=\w^{0,1}=0} = F^+$, where $F^+$
denotes the self-dual part of the standard curvature two-form.
Note that $\cmmib{F}^+$ also contains anti-self-dual
two-form part as well.

Now we examine the equations for fermionic zero-modes.
The equation of motions for fermions are, modulo infinitesimal
gauge transformations,
\eqn\juig{
\eqalign{
i\cmmib{D}^\ppr \bar\eta_- + (\cmmib{D}^\ppr)^* 
\bar\Upsilon^{(0,2)}_- =0,\cr
(\cmmib{D}^\ppr)^* \P^{(0,1)}_+ =0,\cr
\cmmib{D}^\ppr \P^{(0,1)}_+=0.\cr
}
}
Using one of the bosonic localization equation $(\cmmib{D}^\ppr)^2=0$,
we find that the fermionic zero-modes are governed by
the following equations
\eqn\juih{
\cmmib{D}^\ppr \bar\eta_- =0,\qquad
\eqalign{
(\cmmib{D}^\ppr)^* \P^{(0,1)}_+ =0,\cr
\cmmib{D}^\ppr \P^{(0,1)}_+=0,\cr
}\qquad
(\cmmib{D}^\ppr)^* \bar\Upsilon^{(0,2)}_- =0.
}
Thus the fermionic zero-modes are elements of cohomology
group of  the following extended Dolbeault complex
\eqn\juii{
0\rightarrow
\cmmib{S}^{(0,0)}\, \ato{\cmmib{D}^\ppr}\,
\cmmib{S}^{(0,1)}\, \ato{\cmmib{D}^\ppr}\,
\cmmib{S}^{(0,2)}
\rightarrow
0,
}
where
\eqn\juij{
\cmmib{S}^{(0,p)} = \bigoplus_{r+s=p}
\O^{0,r}(M, \wedge^s(T^{*1,0}_M)\otimes End(E)).
}
The net ghost number violation in the
path integral measure due to fermionic zero-modes
is $(\tilde \triangle ,\tilde\triangle)$ where
$\tilde\triangle$ is the negative of the index of the
above complex.
Almost all of the standard procedure in Donaldson-Witten
theory can be repeated here. For example
observables
are $\CG$-equivariant closed differential forms, after the
parity change, on the space $T^*\!\CA$. 
As for a canonical observable we have the $\CG$-equivariant
K\"{a}hler form, after the parity change, on $T^*\!\CA$;
\eqn\juik{
\hat \varpi^\CG_T = \Fr{i}{4\pi^2}\int_M
\tr\biggl( \phi_{++} \cmmib{F}\biggr)\wedge \o
+\Fr{1}{4\pi^2}\int_M\tr\biggl(\P^{(0,1)}_+\wedge\bar\P^{(1,0)}_+\biggr)
\wedge\o,
}
The correlation functions of supersymmetric observables 
are the path integral representations of a generalized
Donaldson-Witten invariant.

We note that the fundamental group of four-manifold does
not seem to play any essential roles in the original Donaldson-Witten
theory. On the other hand the most crucial application
of Simpson's Higgs bundle is on the non-Abelian
Hodge theory associated with the representation
variety $\pi_1(M)\rightarrow GL(r,\msbm{C})$ of
the fundamental group. For this purpose
let us  consider the case the 
$c_1(E)=c_2(E)=0$.\foot{It is not obvious if the moduli space
of stable Higgs bundles has a hyper-K\"{a}hler structure.
For the flat case the existence of hyper-K\"{a}hler
structure has proved by Fujiki \Fujiki.}
It is known that  there is a one-to-one correspondence
between irreducible representations of $\pi_1(M)$ and
stable Higgs bundles with vanishing Chern classes, see 
\SimpsonA.
In this situation Donaldson-Witten invariants concern
only the unitary irreducible representation variety.
An important property of the  moduli space
of stable Higgs bundles is the existence of a $\C^*$ 
action $(E,\w^{1,0})\rightarrow (E, t \w^{1,0})$. 
Simpson showed that  the 
fixed points of $\C^*$ action correspond to complex
variations of Hodge structures. It also implies that any
other representation of $\pi_1(M)$ can be deformed
to a complex variation of Hodge structures. Among the
fixed points the trivial complex variation of Hodge structures
corresponds to unitary irreducible representations.
A useful viewpoint of the $\C^*$ action is to regard
it as Hodge decomposition of non-Abelian cohomology.
Then a unitary representation is some kind of zero-form.
The above results also imply that the path integral of
our model for $c_1(E)=c_2(E)=0$ can be written
as the sum of contributions from every complex variation
of Hodge structures.
Thus it is natural to hope that
our new invariants may have
information beyond the Donaldson-Witten and Seiberg-Witten
invariants for non-simply connected K\"{a}hler $2$-folds.
Of course we do not need to restrict our attention to the
flat case.

The moduli space of stable Higgs bundles have
many beautiful properties and applications.
One of the properties is that the rank $r$ 
stable Higgs sheaves on $M$
can be identified with stable sheaves on the cotangent
bundle $T^*\!M$ which are supported on Lagrangian
subvarieties of $T^*\! M$ which are finite degree $r$
branched coverings of $M$ \SimpsonB\DoMar.
The above property may be relevant to
generalized mirror symmetry on Calabi-Yau $4$-folds \SYZ\Vafa.
If we consider the complex $2$-torus, $T^4$, its cotangent
bundle may be regarded as local model for $T^4$-fibered
Calabi-Yau $4$-folds. Then the moduli space of stable
rank $r$ Higgs sheaves may be viewed as parameterizing
$r$ $D4$-branes wrapped on Lagrangian cycles of Calabi-Yau
$4$-folds. Of course the above picture is too naive but
somewhat suggestive.
Here we will not be able to penetrate many of the applications
and properties of Higgs bundles. 
We will use its $S^1$ symmetry to have an anatomy
of our invariants.

\subsec{Flows to Donaldson-Witten theory}

In the laymen's terms Donaldson-Witten invariant
is simply the symplectic volume of the moduli space
$\CM$ of stable bundles on $M$. Similarly,
the invariants defined by the correlation
function $\left<\exp(\hat\o^\CG_T)\right>$ is the symplectic
volume of the moduli space $\CN$ of stable Higgs
bundles. One of most important properties of the moduli space
$\CN$ is that it has a symmetry under a $S^1$-action, which
can be extended to a $\C^*$-action. The beautiful fact is
that the $\C^*$ action is a very special one, related with
a certain variation of Hodge structures\foot{This
notion will be relevant to the case when the Higgs bundle
is flat. Then $\cmmib{D}=\cmmib{D}^\pr + \cmmib{D}^\ppr$ can be identified
with the Gauss-Manin connections of the associated local system.
Then our localization equations are familiar $tt^*$-equations
in  special geometry \CeVa. In fact for any complex, not necessarily
integral, variation of Hodge structures there is a corresponding
flat Higgs bundle.
}

First we note that our localization equations in \noaga\
are more than the equations \tolb. 
We may replace $\cmmib{D}^\ppr$
by a family of extended derivatives by introducing a spectral
parameter $t$,
\eqn\juil{
\eqalign{
\cmmib{D}^\ppr = \Dpp + t \w^{1,0},\cr
\cmmib{D}^\pr = \Dp + \bar t \w^{0,1}.\cr
}
}
Then our localization equations in \noaga\ imply
that
\eqn\juim{
\eqalign{
(\cmmib{D}^\ppr)^2=0,\cr
\L(\cmmib{D}^\pr\circ \cmmib{D}^\ppr +\cmmib{D}^\ppr\circ \cmmib{D}^\pr)=0,\cr
}
}
for any $t$ with $t \bar t =1$.
Similarly we replace the extended fields defined in \juicy\
as follows
\eqn\wfa{
\eqalign{
\cmmib{A}^{(0,1)}&:= A^{0,1}+ t \w^{1,0},\cr
\Psi^{(0,1)}_+ &:= \p^{0,1}_+ + t \l^{1,0}_+,\cr
\Upsilon_-^{(2,0)} &:= \chi^{2,0}_+ 
+\bar t \chi^{1,1}_- +\bar t^2 \eta^{0,2}_-,\cr
\cmmib{H}^{(2,0)} &:= H^{2,0} + \bar t  H^{1,1} + \bar t^2 K^{0,2},\cr
}\qquad
\eqalign{
\cmmib{A}^{(1,0)} &:= A^{1,0}+\bar t \w^{0,1},\cr
\bar\Psi^{(1,0)}_+ &:= \bar\p^{1,0}_+ + \bar t\bar\l^{0,1}_+,\cr
\bar\Upsilon_-^{(0,2)} &:= \bar\chi^{0,2}_- +  t  \bar\chi^{1,1}_- 
+  t^2 \bar\eta^{2,0}_-,\cr
\cmmib{H}^{(0,2)} &:= H^{0,2} + t \bar H^{ 1,1} + t^2 K^{2,0}.\cr
}
}
Then our action functional $S$ in \slcpp\ or \juid\
is invariant for any $t$ with $t\bar t=1$.

We will show shortly that the $S^1$
action can be extended to a $\C^*$ action by "gauging"
the $U(1)=S^1$ symmetry and scaling the unit $U(1)$ charge.
Such a procedure is equivalent to giving physical bare mass $m$
to the $U(1)$ charged fields. Thus one can consider
an imaginary $\CP^1$ where the $\C^*$ action covers
the natural $\C^*$ action on $\CP^1$ with limit
points $(t=0,t=\infty)$. Now we can identify the
two limit points in $\CP^1$ with
$(m=\infty, m=0)$. Thus we can interpret the absolute
flow generated by the $\C^*$ action as a renormalization
group flow from the past or unbroken phase $m=0$
to the future (present) or broken phase $m\rightarrow \infty$.
This is not just a mere fantasy since we indeed have a
twistor space constructed from the function space of fields
namely the total space $T^*\!\CA$ of the cotangent
bundle over the space of all gauge fields. Our field
space has a hyper-K\"{a}hler structure preserved
by the $\CG$ as well as by the 
$S^1$ symmetry acting on the fiber of $T^*\!\CA$. Such a $S^1$
action can be extended to a $\C^*$ action and then cover
the $\C^*$ action of $\CP^1$ in the twistor
space $T^*\!\CA\times \CP^1$. Furthermore
the Hamiltonian of the $S^1$-action on the field
space is precisely the physical bare mass of the bosonic
fields, whose field space are the fiber of $T^*\CA$ on space-time $M$.
Now by taking the $m\rightarrow \infty$ limit the dominant contributions
to path integral come from the critical 
points of the Hamiltonian, equivalently from the fixed points of $S^1$-action. 
Similarly in the $t\rightarrow 0$
limit any point in the field space flows to a certain fixed point
of the $S^1$-action.  In the trivial fixed point $\w^{1,0}=0$
we recover original Donaldson-Witten theory.
As a global supersymmetric field theory on $M$ 
certain path integral of our model will be localized to a finite dimensional
subspace $\CN$ of the hyper-K\"{a}hler quotient of
$T^*\!\CA$ by $\CG$.
The above argument is valid  regardless whether  $\CN$ preserves
the hyper-K\"{a}hler structure or not.

We may ask an interesting physical question. Donaldson-Witten
theory is the twisted $N=2$ supersymmetric Yang-Mills
theory. On a manifold with trivial canonical line bundle
twisting does nothing and we have space-time supersymmetric
Yang-Mills theory. Then where shall  we place our model?
Our proposal is that it may describe a certain unbroken
phase of bigger symmetry which is connected to the physical
super-Yang-Mills
theory by renormalization group flows, and the physical
theory lives in one of the fixed points.  

Now we  perturbed our model
by "gauging" the $U(1)$ symmetry.
For this we modify the
supersymmetry transformation laws according to
the following anti-commutations relations
\eqn\wfac{
\bs_+^2 =0,\qquad \{\bs_+,\bbs_+\} = -i\phi^a_{++}\CL_a - i m \CL_{S^1},
\qquad \bbs_+^2 =0.
}
We define a new action functional $S(m)$
by the same formula as $S$ in \juid\ but with
the modified transformation laws. Then
we define a family of  $N_c=(2,0)$ models parameterized 
by $m$ and $\bar m$
with the following action functional
\eqn\wfad{
S(m,\bar m) = S(m) + i \bar m \bs_+\bbs_+ \CK(\cmmib{D}),
}
where $\CK(\cmmib{D})$ is the K\"{a}hler potential of $T^*\! \CA$
given by \lcd. Then the action functional contains bare mass
terms for all the charged fields under the $U(1)$, except for auxiliary fields.
The relevant terms
in the action functional looks like
\eqn\wfae{
\eqalign{
S(m,\bar m) &=
S 
\cr
&
-\Fr{\bar m}{4\pi^2}\int_M\tr\left(i\phi_{++} 
(F +[\w^{1,0},\w^{0,1}]) +\p^{0,1}_+\wedge
\bar\p^{1,0}_+ 
+\l^{1,0}_+\!\wedge\!\bar\l^{0,1}_+
\right)\wedge\o
\cr
&
+ \Fr{i m\bar m}{4\pi^2}
\int_M\! \tr\left(\w^{1,0}\!\wedge\!  \w^{0,1}\right)\wedge\o
+\ldots.
}
} 
In the above the $m\bar m$ dependent term is the
Hamiltonian of the $S^1$-action on $T^*\!\CA$.
The term in  the second line
is the equivariant K\"{a}hler
form $\widehat\varpi^\CG_T$ of $T^*\CA$.
Thus  $\widehat\varpi^\CG:=\widetilde\varpi^\CG_T|_\CA$ 
is an observable of
Donaldson-Witten theory which
will descend to the K\"{a}hler form of moduli space $\CM$
of anti-self-dual connections.

Now by taking the $m\rightarrow \infty$ limit we see that
the dominant contributions to the path integral come
from the critical points of the Hamiltonian of the $S^1$-action.
Such critical points are identical to the fixed points of
the $S^1$-action. As usual we always have trivial fixed points
given by $\w^{1,0}=0$ and the fixed point locus is the
moduli space $\CM$ of anti-self-dual connections. 
Thus the contribution from
the trivial fixed points to the partition function
of the model with the action functional $S(m,\bar m)$
is given by a generating functional 
$\left<\exp(\hat\o^\CG)\right>_{DW}$
of  Donaldson-Witten theory weighted by one loop
contributions from the degrees of freedom normal
to $\CM$ in $\CN$. We also note that the value
of the Hamiltonian of the $S^1$-action at the trivial fixed point
is zero.
There are other non-trivial
fixed points $\w^{1,0}\neq 0$ if the $S^1$-action
can be undone by the gauge transformations,
\eqn\wfb{
g \w^{1,0} g^{-1} = t \w^{1,0},
}
where $g \in\CG$ and $t\in U(1)$.

\newsec{Generalized Vafa-Witten Theory}

A class of equivariant
$N_c=(2,0)$ sigma model in zero dimensions
can be extended to a $N_c=(2,2)$
model \ParkB.\foot{A $N_c=(2,2)$ (or $N_c=(2,0)$)
model can be viewed as the dimensional reduction
of $N_{ws}=(2,2)$ (or $N_{ws}=(2,0)$) world-sheet 
supersymmetric gauged
linear sigma model in two-dimensions \GLSM.
In the present case both the target space and the
symmetry group are infinite dimensional. 
Related examples can be found in \HP.
In our terminology a cohomological field theory \Wittencoft\ 
is an equivariant $N_c=(1,0)$ sigma model in zero dimensions,
while a balanced cohomological field theory \DM\
is an equivariant $N_c=(1,1)$ sigma model in zero dimensions.
With a K\"{a}hler structure on the target space the
number of global fermionic symmetry is doubled.  
}
The essential point of such
a construction is introducing additional bosonic fields
corresponding to the local frame fields on the
image of the section $\eufm{S}:\CA\rightarrow \msbm{E}$
so that one has supersymmetric sigma model in zero-dimension
with the target space being the total space of the bundle $\E\rightarrow
\CA$. Then the $\CG$-equivariant holomorphic section $\eufm{S}$
of $\E\rightarrow\CA$ should be extended to gradient vector
of a $\CG$-invariant holomorphic function of the total space
of the bundle $\E\rightarrow\CA$.

Such $N_c=(2,2)$ extension of Donaldson-Witten
theory corresponds to Vafa-Witten theory on a K\"{a}hler surface \DPS. 
Then one can define a family of $N_c=(2,0)$ models by certain
massive perturbation using a natural $S^1$ symmetry of the
$N_c=(2,2)$ model, see also 
\VW\Monad\LaLB.
By taking the bare mass to infinity one see that there are
two different semi-classical limits corresponding to
Donaldson-Witten and Seiberg-Witten theories.
Then the $S$-duality of $N=4$ SYM theory \VW\SWb\
implies the equivalence between Donaldson-Witten and
Seiberg-Witten invariants \DPS\LaLC.

In this section  we apply the above construction
to embed the $N_c=(2,0)$ model in the previous section
a $N_c=(2,2)$ model. The resulting model
generalizes Vafa-Witten theory and compute Euler characteristic
of the moduli space of stable Higgs bundles together
with extra  contributions. 
Then we define $\C^*$
family of  $N_c=(2,2)$ models which has various
limits, which includes  the original Vafa-Witten theory. 
We also perturb the model to $N_c=(2,0)$ model 
and show that the partition function of the theory
is sum of contributions of the generalized Donaldson-Witten
theory in the previous section and a generalized version
of Seiberg-Witten theory. 

\subsec{The $N_c=(2,2)$ Model}

We recall the basic setting for the previous $N_c=(2,0)$
model. We considered the total space $T^*\!\CA$ 
of the  cotangent bundle of the space of all connections
of a rank $r$ Hermitian vector bundle $E\rightarrow M$
over a K\"{a}hler surface $M$. As for the holomorphic
coordinate fields  on $T^*\!\CA$ 
we have the extended connection
$\cmmib{A}^{0,1}$ with superpartner $\P^{0,1}_+$.
We also considered an infinite dimensional $\CG$-equivariant
holomorphic vector bundle $\tilde \msbm{E}\rightarrow T^*\! \CA$
with holomorphic section $\eufm{S}(\cmmib{D}^\ppr) 
= (\cmmib{D}^\ppr)^2:=\cmmib{F}^{(0,2)}$
and associated anti-ghost multiplet 
$(\Upsilon^{(2,0)}_-, \cmmib{H}^{(2,0)})$.

The basic idea behind the extension to a $N_c=(2,2)$
model is that one can regard the total space
of the holomorphic bundle $\tilde\msbm{E}\rightarrow T^*\!\CA$
as the target space of a $N_c=(2,2)$
model. Then we have to supply local holomorphic
coordinate fields for fiber space of  
$\tilde\msbm{E}\rightarrow T^*\!\CA$.
Thus we introduce adjoint-valued bosonic  spectral 
fields $\cmmib{B}^{(2,0)}$ and its superpartner $\Upsilon^{(2,0)}_+$.
Now the former holomorphic section 
$\tilde\eufm{S}=\cmmib{F}^{(0,2)}(\cmmib{D}^\ppr)$
of the bundle $\tilde\msbm{E}\rightarrow T^*\!\CA$ corresponds
to a holomorphic vector field on the target space $\tilde\msbm{E}$
but being supported only on $T^*\!\CA$. Thus the
$\CG$-equivariant holomorphic vector  $\eufm{S}(\cmmib{D}^\ppr)$ 
should be 
extended over the whole space $\tilde\msbm{E}$.
Furthermore  $N_c=(2,2)$ supersymmetry
demands that  a such holomorphic vector should be the
gradient  vector 
of a non-degenerated $\CG$-invariant holomorphic function 
$\CW$, i.e, $\bbs_+\CW=0$,  
on the target space $\tilde\msbm{E}$.

Now demanding $N_c=(2,2)$ supersymmetry will take care 
of everything.
{}From the $N_c=(2,0)$ holomorphic multiplets 
$(\cmmib{A}^{(0,1)},\P^{(0,1)}_+)$ 
we build up
the following chiral multiplets, i.e., $\bbs_\pm \cmmib{A}^{(0,1)}=0$
\eqn\wfd{
\def\normalbaselines{\baselineskip20pt
\lineskip3pt \lineskiplimit3pt}
\matrix{
\P^{(0,1)}_- &\mapl & \cmmib{A}^{(0,1)} &\mapr & \P^{(0,1)}_+\cr
             & 
 	 \rlap{\lower.3ex\hbox{$\scriptstyle s_{\!+}$}}\searrow
 & &\swarrow\!\!\!\rlap{\lower.3ex\hbox{$\scriptstyle s_{\!-}$}} 
& \cr
&&\cmmib{H}^{(0,1)}&&
}.
}
{}From the $N_c=(2,0)$ Fermi multiplets 
$(\Upsilon^{(2,0)}_-, \cmmib{H}^{(2,0)})$ we
build up another set of chiral multiplets, 
i.e., $\bbs_\pm \cmmib{B}^{(2,0)}=0$
\eqn\wfe{
\def\normalbaselines{\baselineskip20pt
\lineskip3pt \lineskiplimit3pt}
\matrix{
\Upsilon^{(2,0)}_- &\mapl & \cmmib{B}^{(2,0)} &\mapr & \Upsilon^{(2,0)}_+\cr
             & 
 	 \rlap{\lower.3ex\hbox{$\scriptstyle s_{\!+}$}}\searrow
 & &\swarrow\!\!\!\rlap{\lower.3ex\hbox{$\scriptstyle s_{\!-}$}} 
& \cr
&&\cmmib{H}^{(2,0)}&&
}.
}
Form the $N_c=(2,0)$ gauge multiplet
$(\phi_{--},\eta_-,\bar\eta_-, H_0,\phi_{++})$
we build up a $N_c=(2,2)$ gauge multiplet
\eqn\wff{
\def\normalbaselines{\baselineskip20pt
\lineskip3pt \lineskiplimit3pt}
\matrix{
\bar\s     & \mapr & \eta_+ & \mapl & \phi_{++}     \cr
\mapd      &       & \mapd  &       & \mapd      \cr
\bar\eta_- & \mapr & H_0      & \mapl & \bar\eta_+ \cr
\mapu      &       & \mapu  &       & \mapu      \cr
\phi_{--}     & \mapr & \eta_- & \mapl & \s         \cr
},
}
which are adjoint valued scalars on $M$.

To keep track of all the fields we write down the
explicit spectral form of the extended fields.
We have
\eqn\wra{
\eqalign{
\cmmib{A}^{(0,1)}&:= A^{0,1}+ t \w^{1,0},\cr
\Psi^{(0,1)}_\pm &:= \p^{0,1}_\pm + t \l^{1,0}_\pm,\cr
\cmmib{H}^{(1,0)} &:= H^{1,0} +  t L^{1,0} ,\cr
}\qquad
\eqalign{
\cmmib{A}^{(1,0)} &:= A^{1,0}+\bar t \w^{0,1},\cr
\bar\Psi^{(1,0)}_\pm &:= \bar\p^{1,0}_\pm + \bar t\bar\l^{0,1}_\pm,\cr
\cmmib{H}^{(0,1)} &:= H^{0,1} + \bar t L^{0,1},\cr
}
}
and
\eqn\wrb{
\eqalign{
\cmmib{B}^{(2,0)} &:= B^{2,0} + \bar t  B^{1,1} + \bar t^2 C^{0,2},\cr
\Upsilon_\pm^{(2,0)} &:= \chi^{2,0}_\pm +\bar t \chi^{1,1}_\pm
+\bar t^2 \eta^{0,2}_\pm,\cr
\cmmib{H}^{(2,0)} &:= H^{2,0} + \bar t  H^{1,1} + \bar t^2 K^{0,2},\cr
}\qquad
\eqalign{
\cmmib{B}^{(0,2)} &:= B^{0,2} + t \bar B^{1,1} + t^2 C^{2,0},\cr
\bar\Upsilon_\pm^{(0,2)} &:= \bar\chi^{0,2}_\pm +  t \bar\chi^{1,1}_\pm
+  t^2 \bar\eta^{2,0}_\pm,\cr
\cmmib{H}^{(0,2)} &:= H^{0,2} + t  \bar H^{1,1} + t^2 K^{2,0}.\cr
}
}

Now we have the standard 
 $N_c=(2,2)$ invariant
functional
\eqn\wfi{
\eqalign{
S=& \bs_+\bbs_+\bs_-\bbs_-\biggl(
\CK(\cmmib{D}) 
+\CK(\cmmib{B}^{(2,0)},\cmmib{B}^{(0,2)})
-\int_M \tr(\s *\bar\s )
\biggr)
\cr
&
+ \bs_+\bs_-\CW\left(\cmmib{A}^{(0,1)}, \cmmib{B}^{(2,0)}\right)
+ \bbs_+\bbs_-\bar\CW\left(\cmmib{A}^{(1,0)}, \cmmib{B}^{(0,2)}\right),
}
}
where
\eqn\wfj{
\CK(\cmmib{B}^{(2,0)},\cmmib{B}^{(0,2)}) 
=-\Fr{1}{4\pi^2} \int_M \tr\left( \cmmib{B}^{(2,0)}\wedge * 
\cmmib{B}^{(0,2)}\right).
}
The holomorphic potential $\CW$, i.e., $\bbs_\pm\CW=0$, 
is given 
as follows
\eqn\wfk{
\eqalign{
\CW\left(\cmmib{A}^{(0,1)}, \cmmib{B}^{(2,0)}\right)&= 
\Fr{1}{4\pi^2}\int_M \tr\left( 
\cmmib{B}^{(2,0)}\wedge * \cmmib{F}^{(0,2)} 
\right).\cr
}
}
We note that the above action functional remains
invariant for any $t$ in \wra\ and \wrb\ with $t\bar t=1$.
We will use this $S^1$ symmetry to define a $\C^*$
family of the $N_c=(2,2)$ model.

Now, from the discussions in Sect.~$3.4$, we see that
the path integral is localized to the zeros of the momentum
map $\m_\msbm{R}$ and the critical points of the holomorphic
potential $\CW$, modulo the gauge symmetry,
\eqn\wfic{
\eqalign{
\cmmib{F}^{(0,2)}=0,\cr
\cmmib{D}^\ppr * \cmmib{B}^{(2,0)}=0,\cr
i\cmmib{F}\wedge \o 
+[\cmmib{B}^{(2,0)}, *\cmmib{B}^{(0,2)}]=0.\cr
}
}
We also have other default localization
equations
\eqn\wfj{
\eqalign{
[\bar\s, \cmmib{B}^{(0,2)}]=0,\cr
[\s, \cmmib{B}^{(0,2)}]=0,\cr
[\s,\bar\s]=0,\cr
\cmmib{D}\bar\s=0,\cr
}\qquad
\eqalign{
[\phi_{\pm\pm}, \cmmib{B}^{(0,2)}]=0,\cr
[\phi_{\pm\pm},\bar\s]=0,\cr
[\phi_{++},\phi_{--}]=0,\cr
\cmmib{D} \phi_{\pm\pm}=0.\cr
}
}
When there are no reducible orbits in \wfic\ we have 
$\s=\phi_{\pm\pm}=0$, and the path integral is localized
to the moduli space defined by \wfic. 
The equation \wfic\ is a generalization of Vafa-Witten
equation. 
We note that the equations in \wfic, as well as in \wfj,
remain the same for any $t$ in \wra\ and \wrb\ with
$t\bar t=1$, which is a symmetry of the action functional.

The equations in \wfic\ have another $S^1$ symmetry
given by 
\eqn\bqa{
(\cmmib{D}^\ppr, \cmmib{B}^{(2,0)})\rightarrow 
(\cmmib{D}^\ppr, \xi\cmmib{B}^{(2,0)})
}
with $\xi\bar\xi=1$. However the above is not
a symmetry of the action functional due to the holomorphic
potential term \wfk;
\eqn\bqb{
S = \Fr{\bs_+\bs_-}{4\pi^2}\int_M
\tr\left(\cmmib{B}^{(2,0)}\wedge *\cmmib{F}^{(0,2)}\right) +\ldots.
}
The above situation is exactly same as for Vafa-Witten
theory on a K\"{a}hler surface. We can use 
the $S^1$ symmetry \bqa\
to break $N_c=(2,2)$ supersymmetry down to $N_c=(2,0)$
supersymmetry by breaking  the supersymmetries 
generated by $\bs_-$ and $\bbs_-$.
We expand \bqb\ by one step to get
\eqn\bqc{
S = \Fr{\bs_+}{8\pi^2}\int_M\tr\left(i\Upsilon^{(2,0)}_-
\wedge  *\cmmib{F}^{(0,2)} 
+ \cmmib{B}^{(2,0)}\wedge * D^\ppr \P^{(0,1)}_-
\right) +\ldots.
}
Then we see that $(\cmmib{D}^\ppr,\cmmib{B}^{(2,0)}, \P^{(0,1)}_-)\rightarrow
(\cmmib{D}^\ppr, \xi\cmmib{B}^{(2,0)}, \bar\xi\P^{(0,1)}_-)$ 
for $\xi\bar\xi=1$
preserves the action functional. On the one hand the
above rotation is not compatible with the supersymmetry
generated by $\bs_-$ since $\bs_- \cmmib{A}^{(0,1)}=i\P^{(0,1)}_-$.
On the other hand we can make it compatible with the
$\bs_+$ supersymmetry by assigning the same $U(1)$
charge to the pair $(\cmmib{B}^{(2,0)}, \Upsilon^{(2,0)}_+)$ 
related by the $\bs_+$ supersymmetry, etc.

In the next section we will use the above $S^1_t\times S^1_\xi$
symmetry to define a $\C^*\times \C^*$ family of
$N_c=(2,0)$ theories. The idea is that all the theories,
both the original and the generalized Donaldson-Witten
and Vafa-Witten theories, we discussed so far should be 
viewed as different semi-classical limits governed by 
different  massless degrees of freedom of the same underlying 
theory.

\subsec{A Family of $N_c=(2,2)$ Models}

We begin with generalizing our $N_c=(2,2)$ model
to a $\C^*$ family of $N_c=(2,2)$ models
using the $S^1_t$ symmetry, whose action is given
as \wra\ and \wrb. For that purpose we extend
our $N_c=(2,2)$ supersymmetry by "gauging"
the $S^1_t$ symmetry;
\eqn\plmd{
\eqalign{
\{\bs_\pm,\bs_\pm\}=0,\cr
\{\bbs_\pm,\bbs_\pm\}=0,\cr
}
\qquad
\eqalign{
\{\bs_+,\bbs_+\} &= -i\phi_{++}^a\CL_a,\cr
\{\bs_+,\bbs_-\}&=-i\s^a\CL_a -i m \CL_{S^1_t},\cr 
\{\bs_-,\bbs_+\} &= -i\bar\s^a\CL_a -i\bar m \CL_{S^1_t},\cr
\{\bs_-,\bbs_-\} &= -i\phi^a_{--}\CL_a,\cr
}
\qquad
\eqalign{
\{\bs_+,\bs_-\}=0,\cr
\{\bbs_+,\bbs_-\}=0,\cr
}
}
where $\CL_{S^1_t}$ denotes the Lie derivative defined
by the vector field generating the $S^1_t$ symmetry.
Equivalently it is an infinitesimal $U(1)$ gauge transformation
for fields with non-vanishing $U(1)$ charge as defined
by \wra\ and \wrb. For convenience we write down explicit
transformations for bosonic fields 
\eqn\simpa{
\eqalign{
\cmmib{A}^{(0,1)}&:= A^{0,1}+ t \w^{1,0},\cr
\cmmib{B}^{(2,0)} &:= B^{2,0} + t^{-1} B^{1,1} + t^{-2} C^{0,2},\cr
}\qquad
\eqalign{
\cmmib{A}^{(1,0)} &:= A^{1,0}+t^{-1} \w^{0,1},\cr
\cmmib{B}^{(0,2)} &:= B^{0,2} + t  \bar B^{1,1} + t^2 C^{2,0}.\cr
}
}

The new action functional $S(m,\bar m)$ is defined by the same formula
as given by \wfi\ but with modified supersymmetry transformation
laws for charged fields under the $S^1_t$.
We write down the relevant terms depending on the bare mass
\eqn\hfes{
\eqalign{
S(m,\bar m) =& S
+\Fr{m\bar m}{4\pi^2}\int_M\!\tr\left( \w^{1,0}\!\wedge \!*\w^{0,1}
+B^{1,1}\!\wedge\! *\bar B^{1,1}
+4 C^{0,2}\!\wedge\! *C^{2,0}\right)
+\ldots
}
}
where the unwritten terms are supersymmetric
completions including the bare mass terms of
$N_c=(2,2)$ superpartners of bosonic fields
charged under $S^1_t$. We remark that the
above action functional preserves all the symmetry
of the original model.
We note that the bare mass terms written above
are exactly the Hamiltonian  of the $S^1_t$ action
on the space of all bosonic fields.
There are two ways of examining the above action
functional. One may take the $|m|\rightarrow \infty$
limit. Then the dominant contributions to the path
integral come from the critical points of the
Hamiltonian of the $S^1_t$ action. Such critical
points are identical to the fixed points of $S^1_t$
action, equivalently the $\msbm{C}^*$ action.
However this viewpoint is rather limited,
as it mainly concerns the moduli space defined
by the equations in \wfic. We should not forget
that such a moduli space is only a subsystem,
and usually  does not form a closed system.

A better viewpoint is to rely on the Higgs
mechanism. We again take the limit that the bare
mass is arbitrarily large. Then we can integrate
out everything except for massless degrees of
freedom. Here the adjoint scalar fields (Higgs fields)
$\s$ and $\bar\s$ play a crucial role since
the effective mass of a field is the sum of the bare
mass and the contribution from the expectation
values of Higgs scalars. This phenomena
can be most directly seen from the  anti-commutation
relations of supercharges \plmd. Since we have
global supersymmetry the expectation
values of supersymmetric observables, $<1> =Z$ in our
case, are localized to an integral over the fixed point
locus of unbroken global supersymmetry.
Consequently the path integral
is localized to the kernel of the right hand sides of $\plmd$
acting on the fields.
Then we immediately get the following set
of relevant equations  for $\cmmib{A}^{0,1}$
and $\cmmib{B}^{2,0}$,
\eqn\simpb{
\eqalign{
[\s,\w^{1,0}] + m\w^{1,0}=0,\cr
[\s,B^{1,1}] - m B^{1,1}=0,\cr
[\s,C^{0,2}] -2 m C^{0,2}=0,\cr
}
}
and
\eqn\simpc{
\eqalign{
\Dpp \s=0,\cr
[\s,\bar\s]=0,\cr
[\s, B^{2,0}]=0,\cr
}
}
We will now study several limits of these equations.

\subsubsection{Three Different Limits}

We consider  an $SU(2)$ bundle $E\rightarrow M$
for simplicity.
The set of equations in \simpb\ are the
conditions for masslessness of the fields charged
under $S^1_t$. The second equation in \simpc\
implies that $\s$ and $\bar\s$ can be
diagonalized, say $\s =\Fr{1}{2}diag(a,-a)$. 
Since $\tr\s^2$ is the gauge invariant object
we will consider $a\geq 0$.

We see that there are three (semi-classical)  limits 
governed by different massless degree of freedom
while preserving $N_c=(2,2)$ supersymmetry.

\begin{enumerate}

\item Vafa-Witten or a twisted $N=4$ super-Yang-Mills
theory. (i) the gauge symmetry is unbroken $a=0$. 
(ii) the gauge symmetry is broken
to $U(1)$ $a > 0$ and $a \neq m, 2m$
Then $\w^{1,0}=B^{1,1}=C^{0,2}=0$ 
is the only solution of \simpb.
Equivalently those fields and their $N_c=(2,2)$ 
superpartners
are all infinitely massive. 

\item The gauge symmetry is broken
to $U(1)$ and $a =m$ and we have the reduction
$E =L\oplus L^{-1}$
Then
\eqn\whua{
\eqalign{
\Dpp & = \left(\matrix{ \bar\rd_{L} & 0 \cr 0 & -\bar\rd_{L}}\right),\cr
\w^{1,0} &= \left(\matrix{0 & 0\cr \vt^{1,0}& 0}\right),\cr
}
\qquad
\eqalign{
B^{2,0} & =  \left(\matrix{ \b^{2,0}& 0 \cr 0 & 
-\b^{2,0}}\right),\cr
B^{1,1} & = \left(\matrix{0 & \b^{1,1} \cr 0& 0}\right),\cr
C^{0,2}&=0.
}
}
We have
\eqn\whub{
\eqalign{
F^{0,2}=0,\cr
i(F -\vt^{1,0}\wedge  \vt^{0,1})\wedge\o +\b^{1,1}\wedge *\bar\b^{1,1}=0,\cr
\bar\rd_{L}\vt^{1,0}=0,\cr
\bar\rd_{L}\b^{2,0} +\vt^{1,0}\wedge *\b^{1,1}=0,\cr
\bar\rd_{L}*\b^{1,1} =0.\cr
}
}

\item The gauge symmetry is broken
to $U(1)$ and $a =2m$ and we have the reduction
$E =L\oplus L^{-1}$
Then
\eqn\whuc{
\eqalign{
\Dpp & = \left(\matrix{ \bar\rd_L & 0 \cr 0 & -\bar\rd_L}\right),\cr
\w^{1,0} &=0,\cr
}\qquad
\eqalign{
B^{2,0} & =  \left(\matrix{ \b^{2,0}& 0 \cr 0 & 
-\b^{2,0}}\right),\cr
B^{1,1}&=0,\cr
C^{0,2} & = \left(\matrix{0 & \g^{0,2} \cr 0& 0}\right).\cr
}
}
We have
\eqn\whud{
\eqalign{
F^{0,2}=0,\cr
\bar\rd_L\b^{2,0}=0,\cr
iF\wedge\o+ \g^{2,0}\wedge \g^{0,2}=0.\cr
}
}

\end{enumerate}

\subsec{Families of $N_c=(2,0)$ Models}

Following the discussions in 
Sect.~$3.4$ and Sect.~$4.2$ 
we break the $N_c=(2,2)$ symmetry
down to $N_c=(2,0)$ supersymmetry
generated by $\bs_+$ and $\bbs_+$.
The $S^1_\xi$-action \bqa\ can be extended
to all those additional
fields introduced for the $N_c=(2,2)$
model, compared with the original $N_c=(2,0)$.
The $S^1_\xi$ action is given by 
\eqn\bdp{
\eqalign{
&S^1_\xi:\left(\cmmib{B}^{(2,0)}, \Upsilon^{(2,0)}_+\right) 
\rightarrow 
\xi\left(\cmmib{B}^{(2,0)},\Upsilon^{(2,0)}_+\right),\cr
&S^1_\xi:\left(\P^{(0,1)}_-, \cmmib{H}^{(0,1)}\right)  
\rightarrow\bar\xi \left(\P^{(0,1)}_-, \cmmib{H}^{(0,1)}\right),\cr
&S^1_\xi:\left(\bar\s,\eta_+\right)\rightarrow 
\bar\xi\left(\bar\s, \eta_+\right),\cr
}
}
and the conjugate fields have the opposite $U(1)_\xi$-charges.
Here we can just follow the procedure in Sect.~$4.2$
to obtain the general $N_c=(2,0)$ supersymmetric
action functional $S(m,\bar m, m_{++},m_{--})$ is given by
\eqn\bdq{
\eqalign{
S(m,\bar m,m_{\pm\pm}) 
=&S(m,\bar m) + m_{++}m_{--}
\int_M\tr\biggl(\cmmib{B}^{(2,0)}\wedge *\cmmib{B}^{(0,2)} 
+ \s*\bar\s \biggr)
+\ldots,
}
}
whose new mass terms contain the Hamiltonian of the $S^1_\xi$
symmetry. The $N_c=(2,0)$ supercharges $\bs_+$ and
$\bbs_+$ satisfy the following modified anti-commutation
relations
\eqn\simma{
\bs_+^2=0,\qquad\{\bs_+,\bbs_+\}= -i\phi_{++}^a\CL_a -
i m_{++}\CL_{S^1_\xi},\qquad \bbs_+^2=0.
}

Now, in total, we have a $\C^*\times \C^*$ 
family of $N_c=(2,0)$ models. From the previous
discussions all we need to do is collect all fixed
point equations of the supercharges $\bs_+$ and
$\bbs_+$. Then the localization equations \wfic\ and
\wfj\ are changed
by the following equations
\eqn\ijkhu{
\eqalign{
\cmmib{F}^{(0,2)} =0,\cr
\cmmib{D}^\ppr *\cmmib{B}^{(2,0)}=0,\cr
\cmmib{F}\wedge \o +[\cmmib{B}^{(2,0)}, *\cmmib{B}^{(0,2)}]
            -\Fr{1}{2}[\s,\bar \s]\o\wedge\o=0,\cr
\cmmib{D}^\pr \s +m\w^{0,1}=0,\cr
[\s ,\cmmib{B}^{(2,0)}] - m B^{1,1} -2m C^{0,2}=0,
}
}
and
\eqn\cson{
\eqalign{
[\phi_{++},\cmmib{B}^{(2,0)}]+m_{++} \cmmib{B}^{2,0} =0,\cr
[\phi_{++},\s]+m_{++} \s=0.\cr
[\phi_{++},\bar\phi_{--}]=0,\cr
d_{\!A}\phi_{++}=0.
}
}
By sending all the bare masses to infinity we have
various semi-classical limits governed by
different massless degrees of freedom.

\subsubsection{Generalized Seiberg-Witten theory}

For our purpose it is suffice to examine  a limit
$m_{\pm\pm}\rightarrow \infty$  by setting $m=\bar m=0$.
For simplicity we consider the $SU(2)$ case.
Then we can follow the  discussions in Sect.~$4.2.2$
and see that the path integral can be written as
the sum of contributions from two branches;

\begin{itemize}

\item branch (i): 
On a generic point on the vacuum moduli space
we have the trivial fixed point $\cmmib{B}^{(0,2)}=0$ and
the fixed point locus is the moduli space $\CN$
of stable Higgs bundles,
\eqn\fsxv{
\eqalign{
\cmmib{F}^{(0,2)}=0,\cr
\cmmib{F}\wedge\o=0.\cr
}
}
Hence we recover the generalization 
Donaldson-Witten theory in Sect. $7.2$.

\item branch (ii): The $SU(2)$ symmetry is broken
down to $U(1)$.
We have $E=L\oplus L^{-1}$
and
\eqn\wfl{
\cmmib{D}^\ppr  =\left(\matrix{\cmmib{d}^\ppr_L & 0 \cr 0 & 
\cmmib{d}^\pr_L}\right),
\qquad
\cmmib{B}^{(2,0)}=\left(\matrix{0 & \cmmib{b}^{(2,0)}\cr 0 & 0}\right),
\qquad
\s=\left(\matrix{0 & \a \cr 0 & 0}\right),
}
where  $\cmmib{d}^\ppr_L = \bar\rd_L + \vt^{1,0}$
and $\cmmib{b}^{(2,0)} = \b^{2,0} +\b^{1,1} + \g^{0,2}$ 
takes values in $L^{-2}$.
The fixed point equation are
\eqn\spsw{
\eqalign{
\cmmib{F}^{(0,2)}&=0,\cr
\cmmib{d}^\pr_L \a &=0,\cr
\cmmib{d}^\ppr_L * \cmmib{b}^{(2,0)} &=0,\cr
i\cmmib{F}_L\wedge\o -\cmmib{b}^{(2,0)}\wedge *\cmmib{b}^{(0,2)}
 + \a \bar\a \o^2&=0.\cr
}
}
The above set of equations is a spectral generalization of  Abelian
Seiberg-Witten equation. For $\th^{1,0}= \b^{1,1}=\g^{0,2}=0$
the above equation reduces to the usual Seiberg-Witten equation
for a special set of Seiberg-Witten classes \VW\DPS.

\end{itemize}

It is a well-established fact that Donaldson-Witten (DW) theory
is equivalent to Seiberg-Witten (SW) theory \SWinv.
One of the strong evidences, or vice versa, for such equivalence
is the $S$-duality  of  Vafa-Witten (VW) theory, which
has both DW and SW theories as two different semi-classical limits
after the massive perturbation. The $S$-duality,
for $SU(2)$ and $SO(3)$, implies that one can recover
the entire partition function from one of such semi-classical limits.
We expect similar relations between the generalized
versions. 
We believe the moduli space of the generalized Seiberg-Witten
equations \spsw\ and associated invariants deserve
detailed study.
It remains to be seen if our invariants contain new information
on smooth structures  beyond Seiberg-Witten
invariants.

\ack

This paper is  based on a chapter of 
the author's PhD thesis defensed by Nov. 1999 in
University of Amsterdam. I would like to
thank my adviser H. Verlinde as well as R. Dijkgraaf and C. Hofman for
valuable discussions. This research is supported by a Pioneer Fund
of NWO and DOE Grant \# DE-FG02-92ER40699.

\end{document}